\documentclass[11pt,a4paper]{article}

\usepackage[utf8]{inputenc}
\usepackage[T1]{fontenc}
\usepackage{geometry}
\usepackage{microtype}
\usepackage{lmodern}

\usepackage{amsmath,amssymb,amsfonts,amsthm}

\usepackage{graphicx}
\usepackage{tikz}
\usetikzlibrary{quantikz2}
\usepackage{subcaption}
\usepackage{booktabs}
\usepackage{authblk}
\usepackage{cite}
\usepackage[colorlinks=true, linkcolor=blue, citecolor=blue, urlcolor=blue]{hyperref}

\title{\textbf{Q-PIPE: A Practical Quantum Phase Encoding Method}}

\author[1]{Brian García Sarmina\thanks{briangarsar12@comunidad.unam.mx}}
\author[2]{Emmanuel Martínez-Guerrero}
\author[2]{Janeth De Anda Gil}
\author[2]{Guo-Hua Sun}
\author[2,3]{Shi-Hai Dong}
\affil[1]{Facultad de Estudios Superiores Cuautitl\'{a}n, Universidad Nacional Aut\'{o}noma de M\'{e}xico, Cuautitl\'{a}n Izcalli, 54714, CDMX, M\'{e}xico}
\affil[2]{Centro de Investigaci\'{o}n en Computaci\'{o}n, Instituto Polit\'{e}cnico Nacional, Gustavo A. Madero, 07700, CDMX, M\'{e}xico}
\affil[3]{Research Center for Quantum Physics, Huzhou Normal University, Huzhou, 313000, China}

\date{\today}

\begin{document}

\maketitle

\begin{abstract}
Efficient loading of classical data into quantum states remains a central bottleneck of quantum computing. We propose Quantum-Gray Phase Injection for Pixel Encoding (Q-PIPE), which writes continuous data into the eigenphases of a diagonal oracle by phase kickback and converts them coherently into computational-basis states by quantum phase estimation. Sums and differences between data sets are obtained by sequential oracle application, without quantum arithmetic circuits. We apply Q-PIPE to quantum image processing, using edge detection by directional finite differences as a proof of concept. Synthesized with uniformly controlled rotations, the oracle requires $O(qN)$ controlled-NOT gates without ancillas for $N$ pixels and $q$ estimation qubits. This is the same scaling as optimized preparation of the novel enhanced quantum representation, whose output form Q-PIPE reproduces. A half-spectrum normalization removes phase aliasing, and a probability threshold derived from the phase-estimation kernel and scaled with image size controls spectral leakage. In ideal simulation, edge detection is exact for inputs commensurate with the estimation register and has low mean absolute error for continuous data. Under device noise models, multi-controlled-phase synthesis drives the output close to the maximally mixed state. The uniformly controlled synthesis instead reduces the compiled depth by a factor of 53--59 and preserves the edge structure (Pearson correlation $r = 0.99$ with the classical gradient). The limiting resource on current hardware is therefore the compiled oracle depth, not the encoding principle.
\end{abstract}


\section{Introduction} \label{sec:intro}

Quantum computing may offer advantages in information processing over classical approaches \cite{Yan2016, khan2024beyond, FAROOQ2025100763}, although data-loading and readout costs can offset them \cite{aaronson2015read}. In imaging, quantum machine learning (QML) has already shown practical benefits, for example in ghost imaging \cite{xiao2023practical, zhai2025quantum}. As hardware matures, the efficient encoding of high-dimensional classical data into quantum states has become a central bottleneck \cite{larose2020robust, weigold2021encoding, khan2024beyond, ZHANG2026114273}.

Several encoding paradigms have been proposed, including basis, amplitude, angle and Hamiltonian encoding. Phase encoding has received comparatively little attention as a standalone data-loading primitive: in QML, data-dependent phases typically serve as feature maps whose information is accessed only indirectly, through kernel estimates or trained observables \cite{havlicek2019supervised, larose2020robust, herrero2026spectral, weigold2020data, weigold2021encoding}, while existing phase-based image representations differ in how the encoded phases are read out and manipulated \cite{du2018synthesis, akinola2024robust}. Here we introduce Quantum-Gray Phase Injection for Pixel Encoding (Q-PIPE), a framework that loads continuous classical data as eigenphases of a diagonal oracle and makes them both computable and readable. It combines three elements: (i) phase kickback from an oracle whose position register is traversed in Gray-code order, which writes every data value into the relative phase of an estimation register; (ii) a single quantum phase estimation (QPE) stage \cite{kitaev1995quantum, cleve1998quantum, NielsenChuang} that coherently converts all encoded phases into computational-basis states, yielding a state of the same form as the novel enhanced quantum representation (NEQR) \cite{Zhang2013}; and (iii) arithmetic between data sets, such as sums and differences, performed directly on the phases before readout, without quantum arithmetic circuits. Aliasing and spectral leakage at readout are controlled by a half-spectrum normalization and a probability threshold derived from the QPE kernel. Items (i) and (ii) together are equivalent to Fourier-space addition \cite{draper2000addition, ruizperez2017quantum}, in which a classical constant is added to a register by phase rotations followed by an inverse quantum Fourier transform ($QFT^{\dagger}$); in Q-PIPE the added constant depends on the position register, which turns the addition into a data loader, and further additions implement item (iii). To the best of our knowledge, Q-PIPE is the first data-loading scheme to exploit this correspondence, combining coherent phase-to-basis readout with native phase-domain arithmetic between data sets. We evaluate it in quantum image processing (QIP), where data-loading costs are particularly severe, using quantum edge detection (QED) as a proof of concept.

Existing QIP representations trade spatial efficiency against operational tractability \cite{rath2024quantum, ranga2024quantum, munikote2024comparing, easom2022efficient, z33f-l4jg}. Amplitude-based representations such as the flexible representation of quantum images (FRQI) \cite{Le2011, ko2025quantum, masum2025quantum, khan2019improved, gonzalez2024efficient} encode pixel intensities into probability amplitudes and require only $O(\log N)$ qubits \cite{leflexible, lu2019multimedia}. Their original preparation circuits require $O(N^2)$ gates, reducible to $O(N)$ with uniformly controlled rotations (UCR) \cite{amankwah2022quantum, mottonen2004quantum}, and the continuous amplitudes, entangled with the position register, complicate subsequent arithmetic and pixel-wise operations.

Basis-encoding schemes such as NEQR \cite{Zhang2013, masum2025quantum, du2022binarization, sang2017novel} store pixel intensities in computational-basis states, enabling direct digital operations such as thresholding and color transformations \cite{sang2017novel, su2021improved}. This simplicity is paid for at initialization: the original multi-controlled preparation scales as $O(qN\log N)$ for $q$-bit intensities \cite{Zhang2013}, reducible to $O(qN)$ with UCR \cite{amankwah2022quantum}.

Frequency- and phase-domain representations have also been explored \cite{lisnichenko2023quantum, majji2023quantum}. Fourier, hyperspectral and cosine-series approaches \cite{grigoryan2020new, alwan2025multilayered, Wereszczynski2020, ko2025quantum} encode images into spectral coefficients, and image data have also been mapped onto phases on the unit circle \cite{grigoryan2025image}. These representations are compact but generally require nontrivial transformations for pixel-wise operations, or costly post-processing to recover spatial information.

Phase-oriented strategies embed pixel values directly into relative phases \cite{du2018synthesis, akinola2024robust, grigoryan2025image}. Their main limitation is readout: phase information is not accessible by projective measurement and typically requires interference-based reconstruction, which increases circuit depth and susceptibility to noise \cite{Werner2023DataLoss}. Later work improved scalability and robustness through circuit optimization and resource reduction \cite{Haque2023EFRQI, amankwah2022quantum}, often at the cost of higher preparation complexity or restricted pixel-level operations.

Section~\ref{sec:pre} presents the Q-PIPE construction and its extension to arithmetic between images. Section~\ref{sec: Results and Simulations} reports the QED benchmarks and the complexity analysis, and section~\ref{sec:discussion} discusses the implications and compares Q-PIPE with existing representations. Section~\ref{sec:conclusions} concludes, and appendices~\ref{app:QED_benchmarks} and \ref{app:noise} give the MNIST reconstructions and the device-noise simulations.


\section{The Q-PIPE framework} \label{sec:pre}

\begin{figure}[t]
  \centering
  \begin{tikzpicture}
    \node[scale=0.95]{
      \begin{quantikz}[row sep=0.65cm, column sep=0.5cm]
        \lstick{$\ket{e_j}$} & \qw & \ctrl{2} & \qw & \ctrl{2} & \qw & \ctrl{2} & \qw & \ctrl{2} & \qw \\
        \lstick{$\ket{p_1}$} & 
          \gate{X} \gategroup[wires=2,steps=1,style={dashed,rounded corners,fill=blue!5},background]{\small Initial map} & 
          \ctrl{1} & 
          \qw \gategroup[wires=2,steps=1,style={dashed,rounded corners,fill=blue!5},background]{\small Transition} & 
          \ctrl{1} & 
          \gate{X} \gategroup[wires=2,steps=1,style={dashed,rounded corners,fill=blue!5},background]{\small Transition} & 
          \ctrl{1} & 
          \qw \gategroup[wires=2,steps=1,style={dashed,rounded corners,fill=blue!5},background]{\small Transition} & 
          \ctrl{1} & 
          \qw \gategroup[wires=2,steps=1,style={dashed,rounded corners,fill=blue!5},background]{\small Uncompute} \\
        \lstick{$\ket{p_0}$} & 
          \gate{X} & 
          \gate[style={fill=cyan!15}]{P(\theta_{00})} & 
          \gate{X} & 
          \gate[style={fill=cyan!15}]{P(\theta_{01})} & 
          \qw & 
          \gate[style={fill=cyan!15}]{P(\theta_{11})} & 
          \gate{X} & 
          \gate[style={fill=cyan!15}]{P(\theta_{10})} & 
          \gate{X}
      \end{quantikz}
    };
  \end{tikzpicture}
  \caption{Gray-code implementation of the oracle $U_{\text{img}}$ for a four-pixel image (position qubits $p_1p_0$), shown for one estimation qubit $e_j$. Each address $x$ is mapped to the all-ones control state by single-qubit $X$ gates, and $P(\theta_x)$ denotes the controlled phase $P(2\pi\,2^j\theta_x)$. Consecutive addresses differ in one bit, so each transition needs a single $X$ gate; the final $X$ gate restores the register, and the block is diagonal in the position basis.}
  \label{fig:qpipe_gray_oracle}
\end{figure}

This section describes the Q-PIPE construction: phase encoding by kickback, readout by QPE through the $\mathrm{QFT}^\dagger$, and the extension to arithmetic between images.

\subsection{Phase encoding construction} \label{sec:phase_enco}

The construction acts on $\mathcal{H} = \mathcal{H}_E \otimes \mathcal{H}_P$, where an $n$-qubit position register $P$ indexes the $N=2^n$ pixel addresses and a $q$-qubit estimation register $E$ supports $2^q$ intensity levels (e.g.\ $q=8$ for 8-bit grayscale). The system starts in

\begin{equation}
\ket{\psi_0} = \ket{0}_E^{\otimes q} \otimes \ket{0}_P^{\otimes n}.
\end{equation}

\textbf{Superposition preparation.} Hadamard gates on both registers give the uniform superposition
\begin{equation}
    \ket{\psi_1} = (H^{\otimes q} \otimes H^{\otimes n})\ket{\psi_0}
    = \frac{1}{\sqrt{2^{n+q}}} \sum_{k=0}^{2^q-1} \sum_{x=0}^{N-1} \ket{k}_E \otimes \ket{x}_P .
    \label{eq:ini_sup}
\end{equation}

\textbf{Phase kickback and encoding.} The central component of Q-PIPE is the image oracle $U_{\mathrm{img}}$, a unitary acting on $\mathcal{H}_P$ that is diagonal in the computational basis \cite{bullock2004asymptotically, welch2014efficient}. Let $I(x)$ be the classical intensity at address $x$ and define the normalized phase $\theta_x = I(x)/L$, where the constant $L$ is fixed by the task. For single-image encoding, $L = 2^{q}$ maps $q$-bit intensities onto $[0,1)$; for arithmetic between images (section~\ref{sec:multi_image}), the half-spectrum choice $L = 2I_{\max}$ confines each phase to $[0,\tfrac{1}{2}]$ (section~\ref{sec: Results and Simulations}). In either case, ideal QPE recovers $\theta_x$ exactly if and only if $2^{q} \cdot \theta_x$ is an integer; otherwise the output is broadened by the finite resolution (section~\ref{res:q_pipe_benchmark_prob_threshold}). The oracle acts as

\begin{equation}
    U_{\mathrm{img}} \ket{x}_P = e^{2\pi i \theta_x} \ket{x}_P ,
    \label{eq:image_oracle}
\end{equation}

so the intensities are its eigenphases. It is implemented with multi-controlled phase gates, and the position register is traversed in Gray-code order (figure~\ref{fig:qpipe_gray_oracle}) \cite{barenco1995elementary, di2021improving, chang2022improving}: consecutive addresses differ in one bit, so each transition requires a single $X$ gate. The estimation qubit $e_j$ controls $U_{\mathrm{img}}^{2^j}$, and together these controlled powers form

\begin{equation}
    \mathcal{U} = \sum_{k=0}^{2^q-1} \ket{k}\!\bra{k}_E \otimes U_{\mathrm{img}}^{k},
\end{equation}

which kicks the phase back onto the estimation register:

\begin{equation}
    \ket{\psi_2} = \mathcal{U}\ket{\psi_1}
    = \frac{1}{\sqrt{N}} \sum_{x=0}^{N-1}
    \left( \frac{1}{\sqrt{2^q}} \sum_{k=0}^{2^q-1} e^{2\pi i k \theta_x} \ket{k}_E \right)
    \otimes \ket{x}_P .
    \label{eq:phase_kickback}
\end{equation}
For each address $x$, the estimation register thus carries the Fourier-basis state of $\theta_x$.

\textbf{Decoding via inverse QFT.} Applying $\mathrm{QFT}^\dagger$ to the estimation register \cite{weinstein2001implementation, camps2021quantum} gives

\begin{equation}
    \ket{\psi_{\mathrm{final}}} = (\mathrm{QFT}^\dagger \otimes I)\ket{\psi_2}
    \approx \frac{1}{\sqrt{N}} \sum_{x=0}^{N-1} \ket{\tilde{I}(x)}_E \otimes \ket{x}_P ,
    \label{eq:qpipe_final}
\end{equation}

where $\ket{\tilde{I}(x)}$ is the $q$-bit basis state nearest to $2^{q} \cdot \theta_x$. Equality holds when $2^q \cdot \theta_x$ is an integer; otherwise the estimation register is peaked around $\ket{\tilde{I}(x)}$ with finite width (section~\ref{res:q_pipe_benchmark_prob_threshold}). The output (figure~\ref{fig:qpipe_single_encoding}) entangles each address with its recovered intensity and therefore has the NEQR form. Unlike in NEQR, the intensities pass through the phase domain before readout, which enables the arithmetic of section~\ref{sec:multi_image}.

\subsection{Arithmetic between images} \label{sec:multi_image}

Because phases add under composition, Q-PIPE extends to arithmetic between images on the same registers, without intermediate measurement. For two images $A$ and $B$ with phases $\theta_{A,x}$ and $\theta_{B,x}$, the product of their oracles acts as

\begin{equation}
    U_{\mathrm{img},B}\,U_{\mathrm{img},A}\ket{x}_P = e^{2\pi i(\theta_{A,x}+\theta_{B,x})}\ket{x}_P .
    \label{eq:utotal_basis}
\end{equation}

Controlled by the same estimation qubits, the two oracle sequences combine into

\begin{equation}
    \mathcal{U}_{\mathrm{total}} = \prod_{m \in \{A, B\}} \sum_{k=0}^{2^q-1} \ket{k}\!\bra{k}_E \otimes U_{\mathrm{img},m}^{k}
    = \sum_{k=0}^{2^q-1} \ket{k}\!\bra{k}_E \otimes \left(U_{\mathrm{img},B}\,U_{\mathrm{img},A}\right)^{k},
    \label{eq:utotal_controlled}
\end{equation}

where the second equality holds because diagonal operators commute. Acting on $\ket{\psi_1}$, this gives

\begin{equation}
    \ket{\psi_3} = \frac{1}{\sqrt{N}} \sum_{x=0}^{N-1}
    \left( \frac{1}{\sqrt{2^q}} \sum_{k=0}^{2^q-1} e^{2\pi i k (\theta_{A,x} + \theta_{B,x})} \ket{k}_E \right) \otimes \ket{x}_P ,
\end{equation}

and the $\mathrm{QFT}^\dagger$ converts the accumulated phase into the computational basis,

\begin{equation}
    \ket{\psi_{\mathrm{final}}} = (\mathrm{QFT}^\dagger \otimes I)\ket{\psi_3}
    \approx \frac{1}{\sqrt{N}} \sum_{x=0}^{N-1} \ket{\tilde{I}_{A+B}(x)}_E \otimes \ket{x}_P ,
\end{equation}

where $\ket{\tilde{I}_{A+B}(x)}$ is the $q$-bit basis state nearest to $2^q(\theta_{A,x}+\theta_{B,x}) \bmod 2^q$. Measuring the estimation register at address $x$ therefore yields an outcome $k$ with

\begin{equation}
    \frac{k}{2^{q}} \approx \theta_{A,x} + \theta_{B,x} \pmod 1 .
\end{equation}

Since phases are defined modulo $1$, the result is unambiguous only if it remains within one period. For addition, the half-spectrum mapping $\theta_x \in [0,\tfrac{1}{2}]$ ($L = 2I_{\max}$) gives $\theta_{A,x}+\theta_{B,x} \in [0,1]$, which is ambiguous only when both intensities equal $I_{\max}$.

\textbf{Subtraction.} Because $U_{\mathrm{img},B}$ is diagonal, its inverse conjugates each eigenphase,

\begin{equation}
    U_{\mathrm{img},B}^{\dagger}\ket{x}_P = e^{-2\pi i\theta_{B,x}}\ket{x}_P ,
    \label{eq:inverse_oracle}
\end{equation}

and replacing $U_{\mathrm{img},B}$ with $U_{\mathrm{img},B}^{\dagger}$ in \eqref{eq:utotal_controlled} kicks back the phase $e^{2\pi i k(\theta_{A,x}-\theta_{B,x})}$. At the circuit level, the inverse oracle keeps the Gray-code structure of figure~\ref{fig:qpipe_gray_oracle}, with each $P(2\pi\, 2^{j}\theta_{B,x})$ replaced by $P(-2\pi\, 2^{j}\theta_{B,x})$; equivalently, the phases of image $B$ are negated classically, $\theta_{B,x}\to-\theta_{B,x}$, before encoding. No ancilla qubits, arithmetic circuits or additional gates are required, subtraction costs exactly the same as addition. With the half-spectrum mapping, $\Delta\theta_x = \theta_{A,x}-\theta_{B,x} \in [-\tfrac{1}{2},\tfrac{1}{2}]$, and the sign is recovered by the decoding rule \eqref{eq:decoding} of section~\ref{sec: Results and Simulations}. Both adjustments are classical, applied at encoding and readout, and leave the circuit unchanged.

\begin{figure*}[t!]
  \centering
  \begin{subfigure}[b]{0.45\textwidth}
    \centering
    \begin{tikzpicture}
      \node[scale=0.53]{
        \begin{quantikz}[row sep=0.5cm, column sep=0.6cm]
          \lstick{$\ket{0}_{e_2}$} & \gate{H} & \ctrl{3} & \qw & \qw & \gate[wires=3]{\mathrm{QFT}^{\dagger}} & \meter{} \\
          \lstick{$\ket{0}_{e_1}$} & \gate{H} & \qw & \ctrl{2} & \qw & & \meter{} \\
          \lstick{$\ket{0}_{e_0}$} & \gate{H} & \qw & \qw & \ctrl{1} & & \meter{} \\
          \lstick{$\ket{0}_{p_2}$} & \gate{H} & 
            \gate[wires=3]{U(\theta)^4} 
            \gategroup[wires=3,steps=3,
              style={rounded corners,fill=cyan!15,inner xsep=2pt,inner ysep=2pt},
              background, label style={label position=below, anchor=north, yshift=-0.2cm}]{\Large Pixel intensity ($\theta$)} & 
            \gate[wires=3]{U(\theta)^2} & 
            \gate[wires=3]{U(\theta)^1} & \qw & \meter{} \\
          \lstick{$\ket{0}_{p_1}$} & \gate{H} & & & & \qw & \meter{} \\
          \lstick{$\ket{0}_{p_0}$} & \gate{H} & & & & \qw & \meter{}
        \end{quantikz}
      };
    \end{tikzpicture}
    \caption{Single-image encoding. The position register $p_{2},p_{1},p_{0}$ indexes the pixels; the estimation qubit $e_j$ controls $U(\theta)^{2^j}$ and receives the kicked-back phase.}
    \label{fig:qpipe_single_encoding}
  \end{subfigure}
  \hfill
  \begin{subfigure}[b]{0.52\textwidth}
    \centering
    \begin{tikzpicture}
      \node[scale=0.50]{
        \begin{quantikz}[row sep=0.7cm, column sep=0.45cm]
          \lstick{$\ket{0}_{e_2}$} & \gate{H} & \ctrl{3} & \qw & \qw & \ctrl{3} & \qw & \qw & \gate[wires=3]{\mathrm{QFT}^{\dagger}} & \meter{} \\
          \lstick{$\ket{0}_{e_1}$} & \gate{H} & \qw & \ctrl{2} & \qw & \qw & \ctrl{2} & \qw & & \meter{} \\
          \lstick{$\ket{0}_{e_0}$} & \gate{H} & \qw & \qw & \ctrl{1} & \qw & \qw & \ctrl{1} & & \meter{} \\
          \lstick{$\ket{0}_{p_2}$} & \gate{H} & 
            \gate[wires=3]{U(\theta_A)^4} 
            \gategroup[wires=3,steps=3,
              style={rounded corners,fill=cyan!15,inner xsep=2pt,inner ysep=2pt},
              background, label style={label position=below, anchor=north, yshift=-0.2cm}]{\Large Image A} & 
            \gate[wires=3]{U(\theta_A)^2} & 
            \gate[wires=3]{U(\theta_A)^1} & 
            \gate[wires=3]{U(\theta_B)^4} 
            \gategroup[wires=3,steps=3,
              style={rounded corners,fill=orange!15,inner xsep=2pt,inner ysep=2pt},
              background, label style={label position=below, anchor=north, yshift=-0.2cm}]{\Large Image B} & 
            \gate[wires=3]{U(\theta_B)^2} & 
            \gate[wires=3]{U(\theta_B)^1} & \qw & \meter{} \\
          \lstick{$\ket{0}_{p_1}$} & \gate{H} & & & & & & & \qw & \meter{} \\
          \lstick{$\ket{0}_{p_0}$} & \gate{H} & & & & & & & \qw & \meter{}
        \end{quantikz}
      };
    \end{tikzpicture}
    \caption{Arithmetic between two images. The controlled oracles of images A and B add their phases, $\theta_A+\theta_B$, before the $\mathrm{QFT}^\dagger$; subtraction uses $\theta_B \to -\theta_B$ (section~\ref{sec:multi_image}).}
    \label{fig:qpipe_phase_accumulation}
  \end{subfigure}
  \caption{Q-PIPE circuits with three position qubits and three estimation qubits.}
  \label{fig:qpipe_circuits}
\end{figure*}

Figure~\ref{fig:qpipe_phase_accumulation} draws the two oracles as separate stages for clarity. Because both are diagonal in the same basis, the gates for address $x$ can be applied at the same Gray-code step and, having identical controls and targets, merged into a single $C^{n}P\big(2\pi\, 2^{j}(\theta_{A,x}-\theta_{B,x})\big)$ gate.


\section{Numerical demonstrations and results} \label{sec: Results and Simulations}

We evaluate Q-PIPE on QED \cite{llorens2025quantum, zhou2020metasurface}, benchmarking it against classical finite differences such as $G_{y}(x, y) = I(x, y) - I(x, y - 1)$ \cite{yao2017quantum, zhang2015qsobel}. Pixel coordinates are written $(x,y)$, with $x$ the row and $y$ the column, and the position register holds the address $\ket{x,y}_P$; $G_y$ is therefore the horizontal difference. The gradient is computed with the subtraction of section~\ref{sec:multi_image}. The original image is encoded through $U_{\mathrm{img},A}$ with

\begin{equation}
    \theta_{A}(x,y) = \frac{I(x,y)}{2I_{\max}},
    \label{eq:pixel_int_enc}
\end{equation}

and the displaced image through the inverse oracle \eqref{eq:inverse_oracle} with $\theta_{B}(x,y) = I(x,y-1)/(2I_{\max})$, i.e.\ $L=2I_{\max}$, where $I_{\max}=255$ for 8-bit images and $I(x,-1)\equiv 0$ at the boundary. At each Gray-code step, estimation qubit $e_j$ controls $P\big(2^{j}\pi I(x,y)/I_{\max}\big)$ and $P\big(-2^{j}\pi I(x,y-1)/I_{\max}\big)$, or their merged gate. The estimation register accumulates the phase difference without intermediate measurement:

\begin{equation}
    \ket{\psi_{\mathrm{QED}}} = \frac{1}{\sqrt{2^{q+n}}} \sum_{x,y} \left( \sum_{k=0}^{2^{q}-1} e^{2\pi i k \Delta\theta(x,y)} \ket{k}_{E} \right) \otimes \ket{x,y}_{P},
    \label{eq:qed_qpipe}
\end{equation}

with

\begin{equation}
    \Delta\theta(x,y) = \theta_{A}(x,y)-\theta_{B}(x,y)
    = \frac{I(x,y)-I(x,y-1)}{2I_{\max}} \in \left[-\tfrac{1}{2},\tfrac{1}{2}\right].
    \label{eq:delta_theta}
\end{equation}

Applying $\mathrm{QFT}^\dagger$ to $E$ then maps all gradient values onto the computational basis at once.

The factor $\tfrac12$ in \eqref{eq:pixel_int_enc} prevents aliasing. Mapping 8-bit intensities onto the full range $[0,2\pi)$ gives phase differences in $(-2\pi,2\pi)$; because phases are $2\pi$-periodic, any difference beyond $\pm\pi$ wraps around, as in classical phase unwrapping \cite{spagnolini19932, zhu2022anti}. 

\begin{figure*}[t!]
    \centering
    \includegraphics[width=\textwidth]{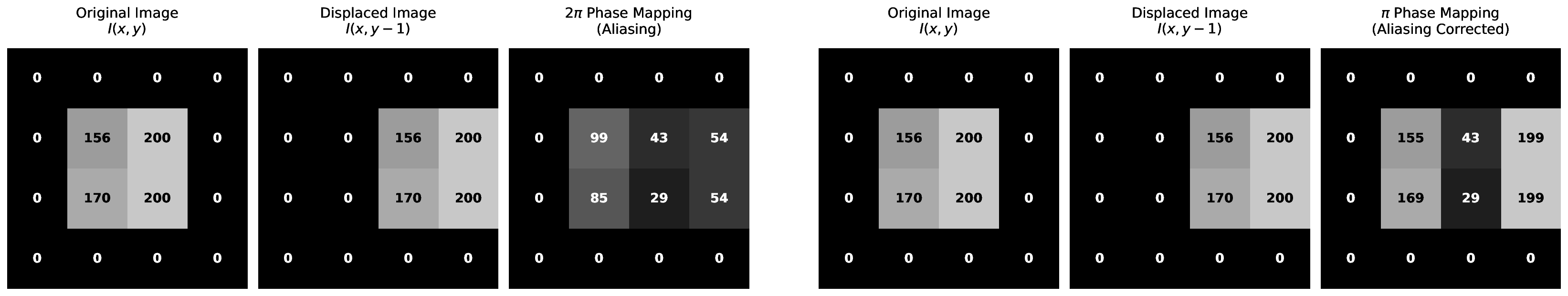}
    \caption{Toy $4 \times 4$ example of QED with Q-PIPE. \textbf{Left:} mapping intensities onto the full phase range $[0, 2\pi)$ makes large gradients wrap around the phase circle (aliasing). \textbf{Right:} mapping each image onto the half-spectrum $[0, \pi]$ and rescaling the output by $2I_{\max}$ removes the wrap-around and recovers the edge magnitudes to within one intensity level.}
    \label{fig:qed_example}
\end{figure*}

In a horizontally shifted $4\times4$ toy image, a discontinuity $\Delta I = -200$ wraps in this way, QPE returns only the shortest cyclic distance, and the recovered magnitudes are wrong (figure~\ref{fig:qed_example}, left). With the half-spectrum mapping, $\theta\in[0,\tfrac{1}{2}]$, the difference \eqref{eq:delta_theta} stays within one period. The outcome $k$ gives $f \equiv k/2^{q} \approx \Delta\theta \pmod 1$, from which the signed gradient and its magnitude are recovered as

\begin{equation}
    \widehat{\Delta I} = 2I_{\max} \times 
    \begin{cases}
        f, & f < \tfrac{1}{2},\\
        f-1, & f \geq \tfrac{1}{2},
    \end{cases}
    \qquad
    |\widehat{\Delta I}| = 2I_{\max}\min(f,\,1-f).
    \label{eq:decoding}
\end{equation}

The factor $2I_{\max}$ undoes the normalization. The sign is ambiguous only when $|\Delta I| = I_{\max}$, where $\Delta\theta=\pm\tfrac{1}{2}$ coincide on the unit circle; the magnitude is unaffected. Figure \ref{fig:qed_example} (right) now shows that the wrap-around is removed.

Because $2^q\Delta\theta$ is generally not an integer, QPE spreads each value over neighboring basis states (spectral leakage) \cite{breitenbach1999against, harris1978use}. Each $|\widehat{\Delta I}|$ is therefore estimated as the probability-weighted mean of \eqref{eq:decoding} over the outcomes $k$ with $P(k,x,y) \geq P_{\mathrm{th}}$, normalized over the retained outcomes, with $P_{\mathrm{th}} = 5\times10^{-4}$ in the scalability benchmarks and $10^{-3}$ in the MNIST demonstrations; section~\ref{res:q_pipe_benchmark_prob_threshold} derives how $P_{\mathrm{th}}$ must scale with $n$. Since only magnitudes are used below, all reported results employ $|\widehat{\Delta I}|$. The vertical and horizontal components, $|G_x|_{\mathrm{QED}}$ and $|G_y|_{\mathrm{QED}}$, obtained from \eqref{eq:decoding} with the image displaced along $x$ and $y$, are combined into the gradient magnitude

\begin{equation}
    M_{\mathrm{QED}} = \sqrt{|G_x|_{\mathrm{QED}}^2 + |G_y|_{\mathrm{QED}}^2},
    \label{eq:sobel}
\end{equation}

as in the Sobel operator but with first-order differences \cite{gonzalez2018digital}. Accuracy is quantified by the mean absolute error (MAE) between the Q-PIPE output $G_{\mathrm{quantum}}$ and the classical reference $G_{\mathrm{classic}}$, averaged over the $N$ pixels.


\subsection{Estimation-register depth} \label{res:q_pipe_benchmark}

\begin{figure*}[t!]
    \centering
    \includegraphics[width=\textwidth]{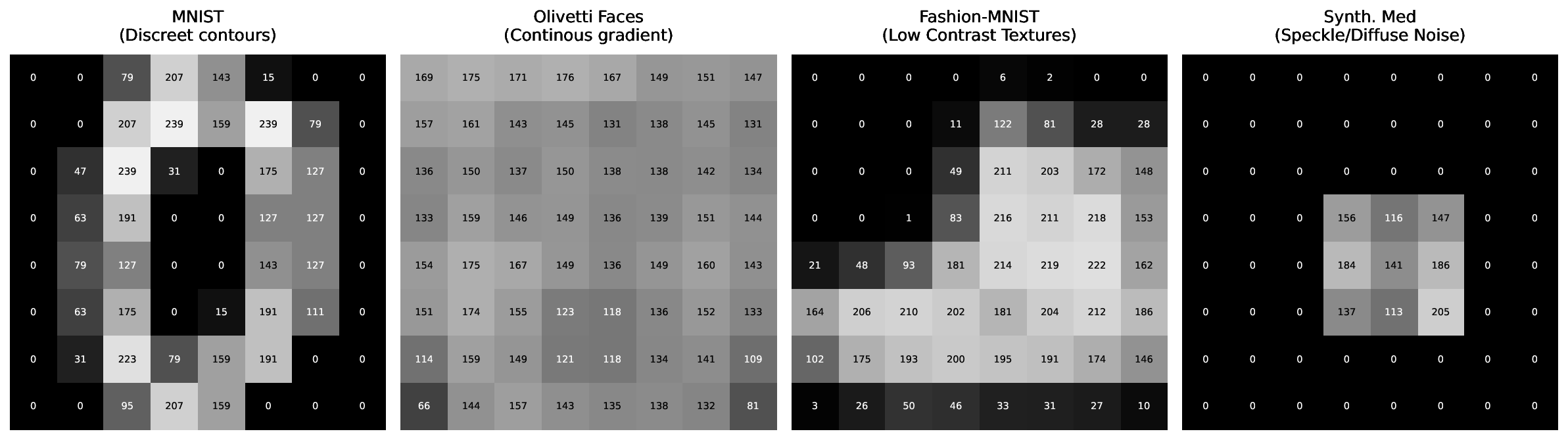}
    \caption{Representative samples from the image datasets used for the performance evaluation of Q-PIPE: MNIST, Olivetti Faces, Fashion-MNIST, and a synthetic medical dataset.}
    \label{fig:q_pipe_benchmark_examples}
\end{figure*}

\begin{figure}[t!]
    \centering
    \includegraphics[height=7cm, width=\textwidth]{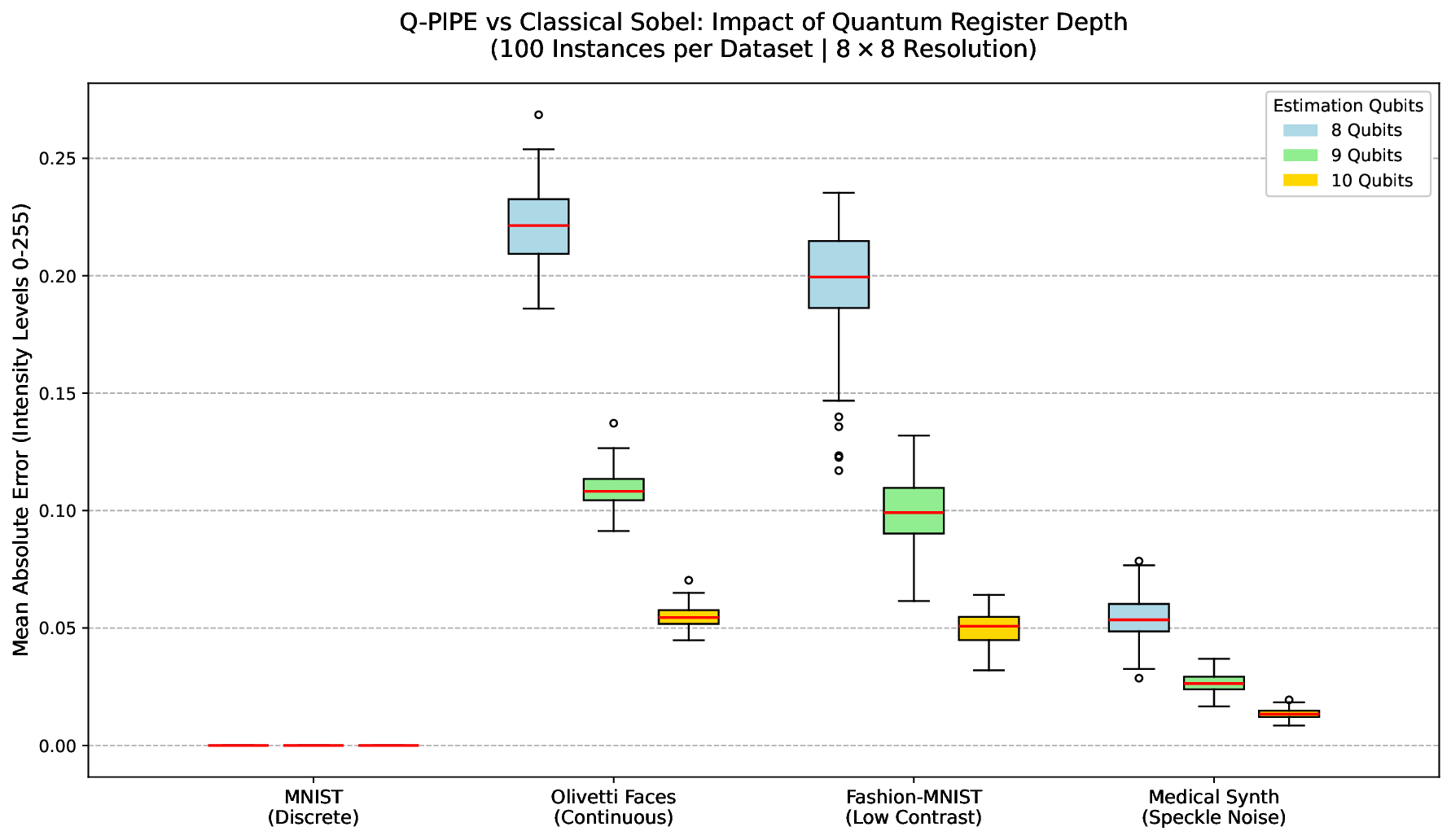}
    \caption{MAE of the gradient magnitude (lower is better) at $8\times8$ resolution for 100 samples per dataset (MNIST: \textit{load\_digits}), with 8 (blue), 9 (green) and 10 (yellow) estimation qubits.}
    \label{fig:q_pipe_benchmark_gruped}
\end{figure}

We evaluate QED on four datasets with distinct spatial characteristics (figure~\ref{fig:q_pipe_benchmark_examples}), all at $8\times8$ resolution: MNIST (discrete contours; the \textit{load\_digits} variant), Olivetti Faces (continuous gradients), Fashion-MNIST (low-contrast textures) and a synthetic medical dataset with speckle noise. For 100 samples per dataset, we compute the MAE of the gradient magnitude, on the $0$--$255$ intensity scale, for estimation registers of 8 to 10 qubits (figure \ref{fig:q_pipe_benchmark_gruped}).

MNIST gives an MAE of exactly zero for every register size: its intensities take only 17 levels, so $2^q\Delta\theta$ is an integer and QPE is exact (appendix~\ref{app:QED_benchmarks}, figures~\ref{fig:s1}--\ref{fig:s3}). For the continuous datasets, $2^q\Delta\theta$ is generally not an integer; the MAE at 8 qubits remains below $0.27$, and the median MAE halves with each additional estimation qubit (e.g.\ $0.22$, $0.11$ and $0.054$ for Olivetti Faces at 8, 9 and 10 qubits), as expected from the QPE resolution $2^{-q}$. At 10 qubits the MAE lies in $0.045$--$0.065$ for Olivetti Faces and $0.008$--$0.019$ for the synthetic medical dataset.

\begin{figure*}[t!]
     \centering
     \begin{subfigure}[b]{0.49\textwidth}
         \centering
         \includegraphics[width=\textwidth]{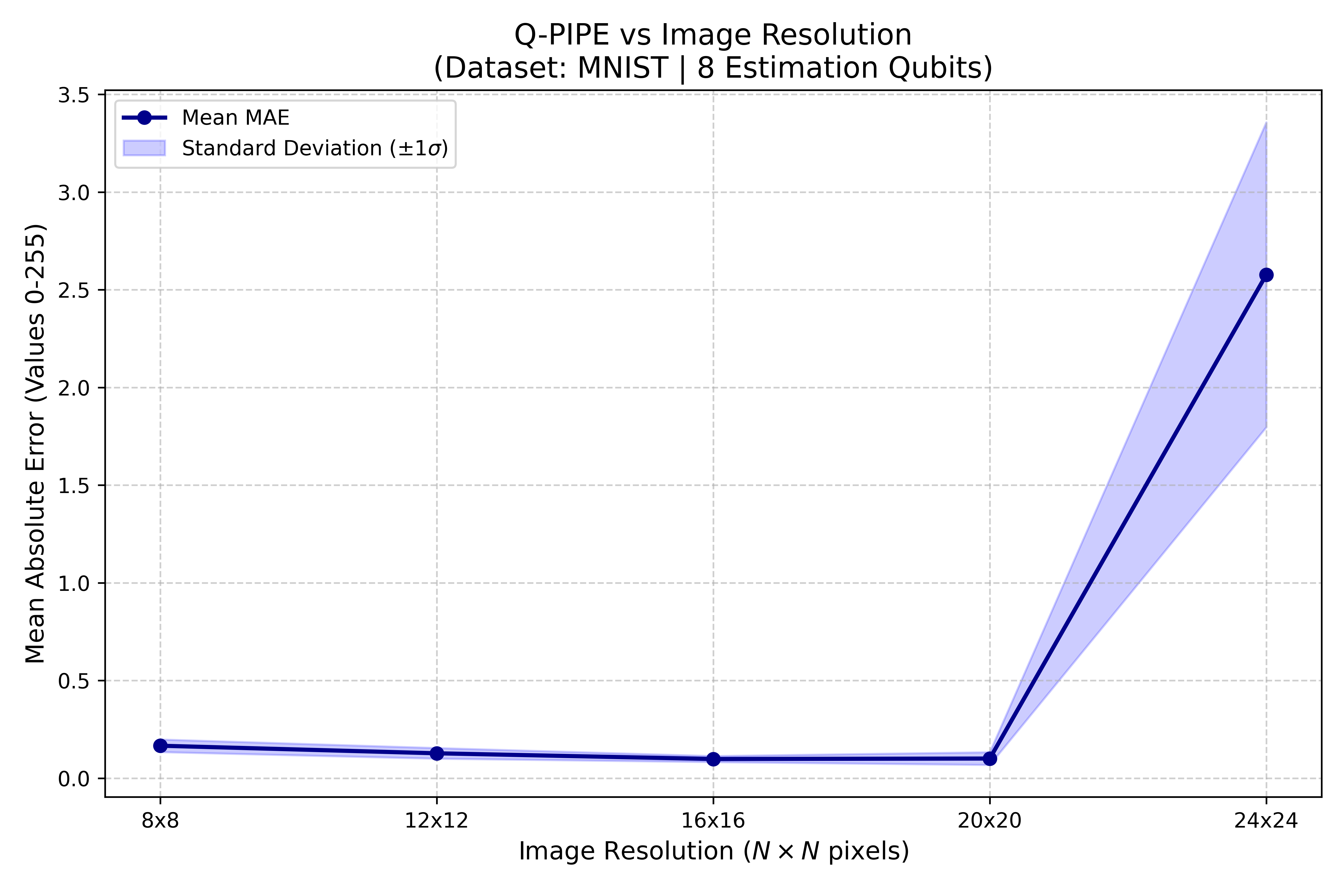}
         \caption{MNIST (\textit{fetch\_openml}) dataset.}
         \label{fig:q_pipe_benchmark_mnist_diff_res}
     \end{subfigure}
     \hfill
     \begin{subfigure}[b]{0.49\textwidth}
         \centering
         \includegraphics[width=\textwidth]{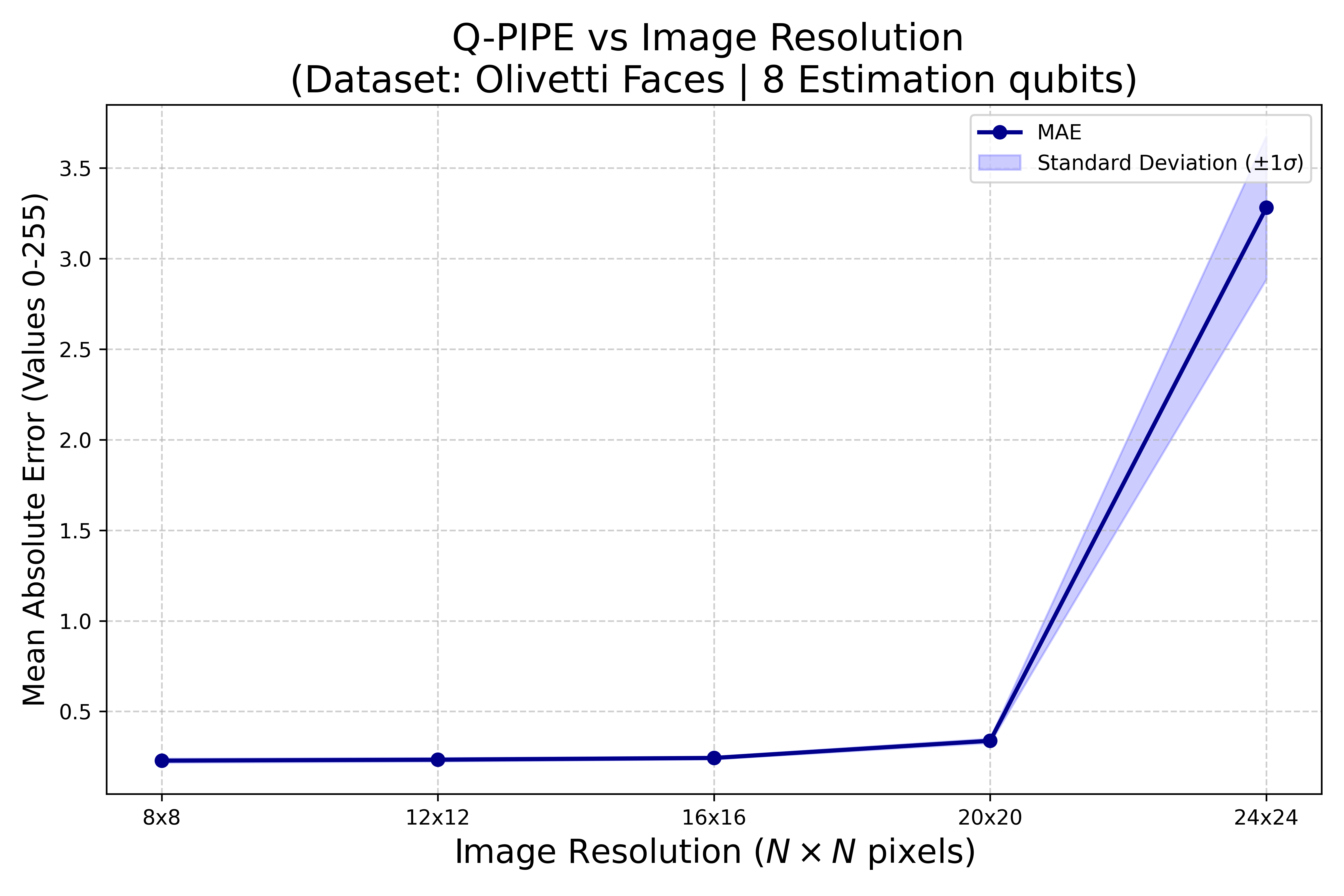}
         \caption{Olivetti Faces dataset.}
         \label{fig:q_pipe_benchmark_olivetti_diff_res}
     \end{subfigure}
     \begin{subfigure}[b]{0.49\textwidth}
         \centering
         \includegraphics[width=\textwidth]{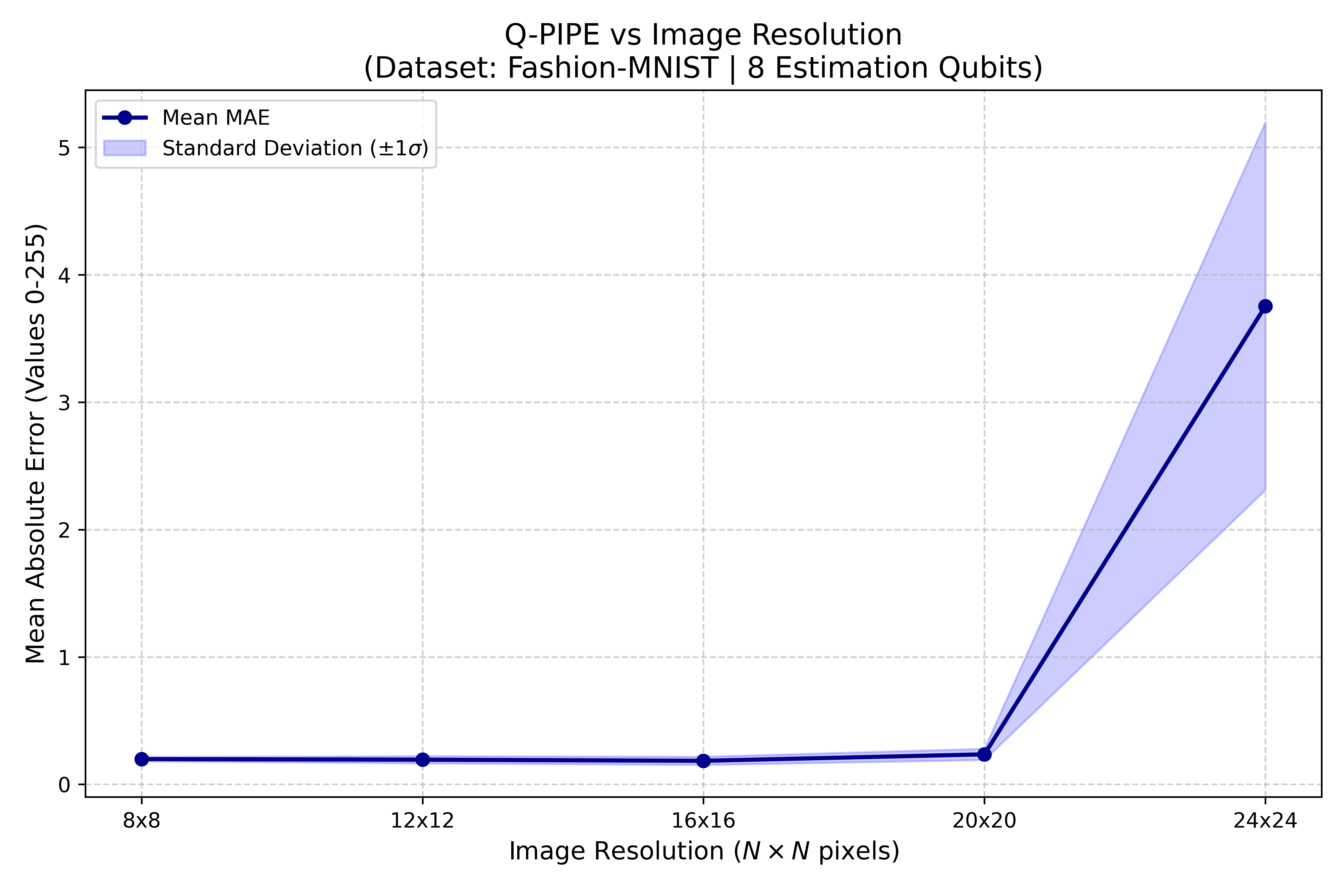}
         \caption{Fashion-MNIST dataset.}
         \label{fig:q_pipe_benchmark_fmnist_diff_res}
     \end{subfigure}
     \hfill
     \begin{subfigure}[b]{0.49\textwidth}
         \centering
         \includegraphics[width=\textwidth]{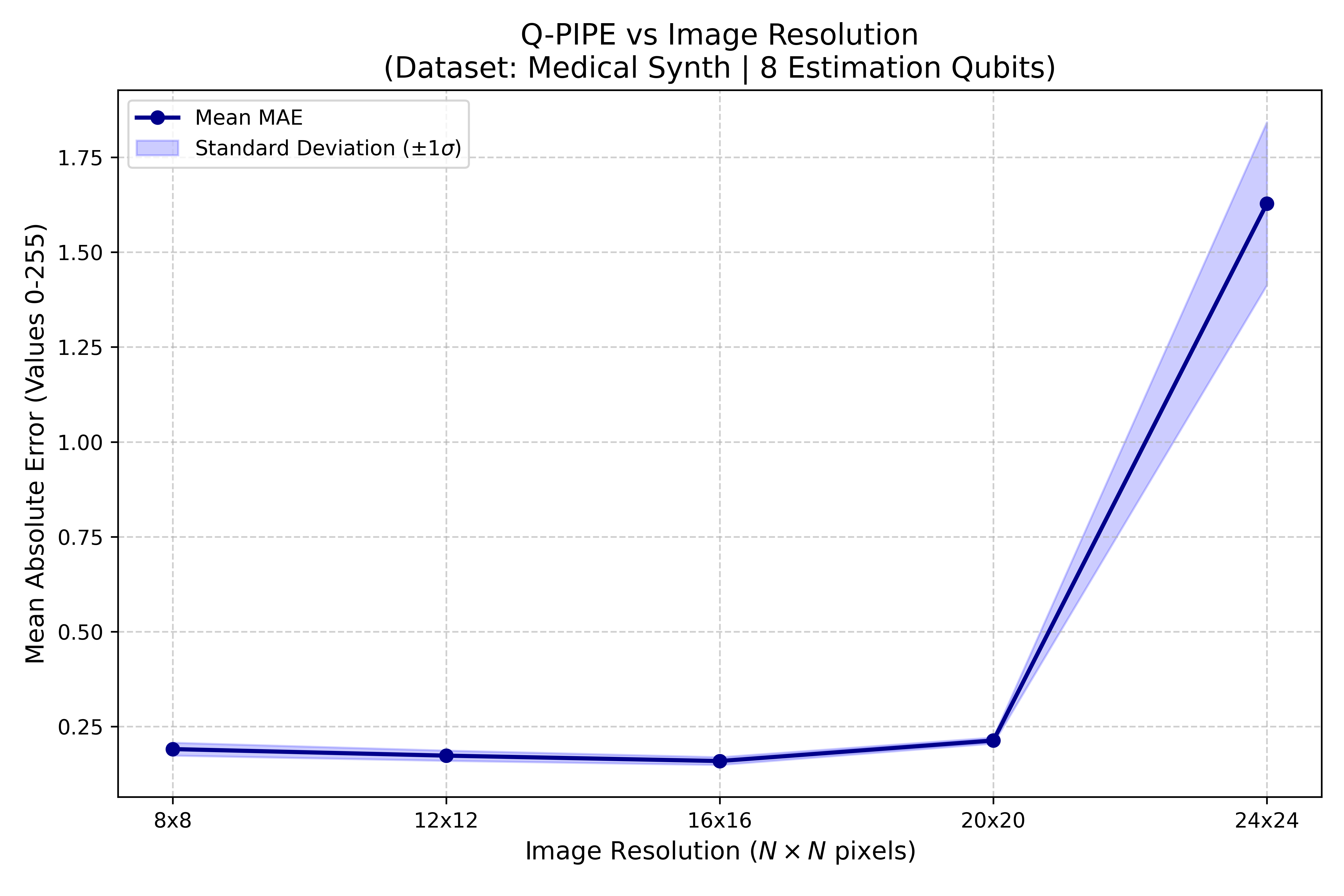}
         \caption{Synthetic medical dataset.}
         \label{fig:q_pipe_benchmark_synmed_diff_res}
     \end{subfigure}
    \caption{MAE of the gradient magnitude (lower is better) versus spatial resolution ($8\times8$ to $24\times24$ pixels) with a fixed 8-qubit estimation register and $P_{\mathrm{th}}=5\times10^{-4}$.}
    \label{fig:q_pipe_benchmark_different_image_resolutions}
\end{figure*}

\subsection{Image resolution} \label{res:q_pipe_benchmark_resolution}

We next vary the resolution from $8\times8$ to $24\times24$ pixels with $q=8$ and $P_{\mathrm{th}}=5\times10^{-4}$, averaging over 10 samples per dataset because the cost of classical simulation grows exponentially (figure~\ref{fig:q_pipe_benchmark_different_image_resolutions}; MNIST here is the \textit{fetch\_openml} variant). Up to $20\times20$ the MAE stays at $\approx 0.09$--$0.35$ for all datasets. At $24\times24$ the mean MAE rises to $\approx 2.6$ (MNIST), $3.3$ (Olivetti Faces), $3.8$ (Fashion-MNIST) and $1.6$ (synthetic medical dataset).

\begin{figure*}[tp]
     \centering
     \begin{subfigure}[b]{0.49\textwidth}
         \centering
         \includegraphics[width=\textwidth]{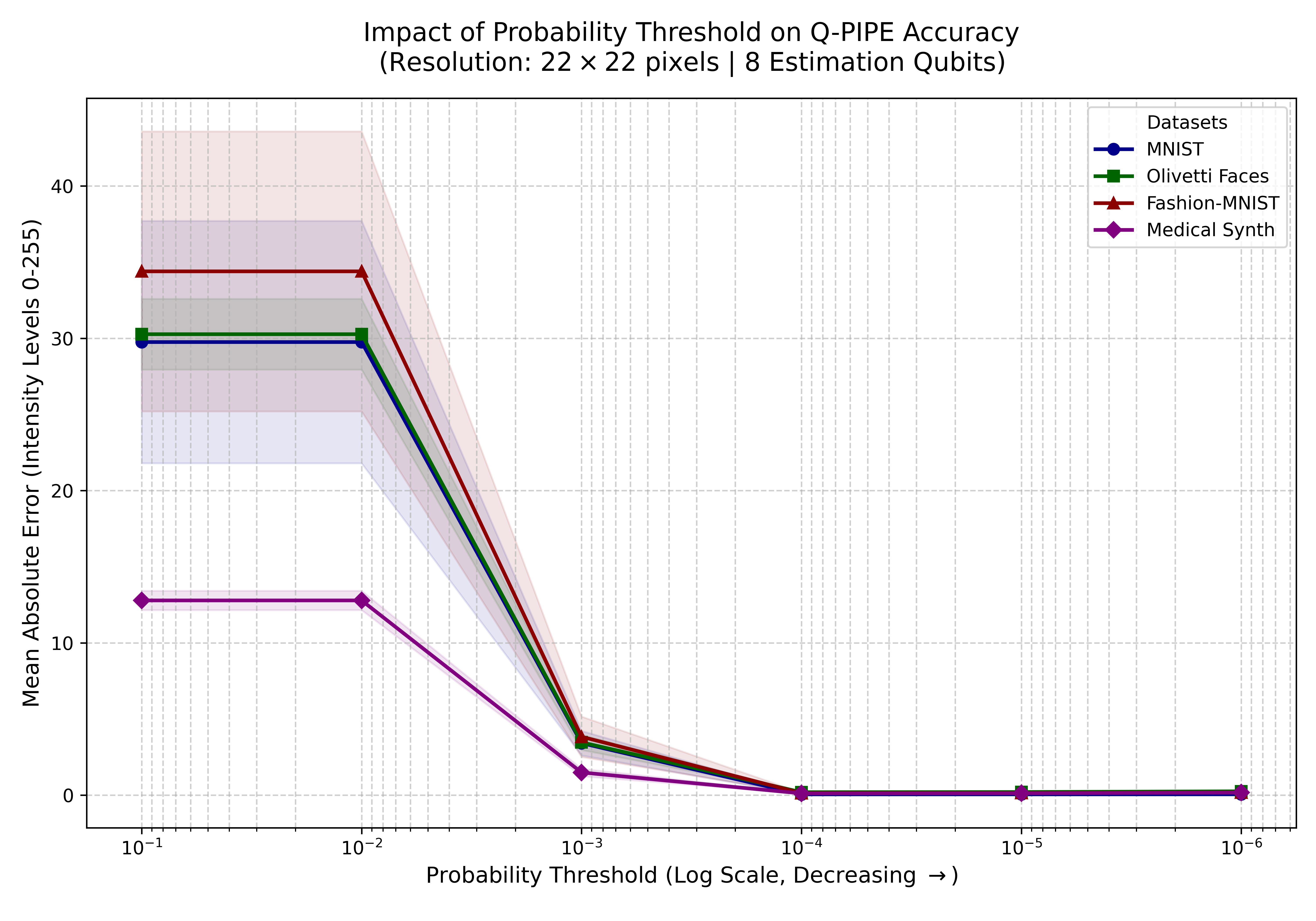}
         \caption{Spatial resolution of $22 \times 22$ pixels.}
         \label{fig:q_pipe_diff_prob_thresholds_22x22}
     \end{subfigure}
     \hfill
     \begin{subfigure}[b]{0.49\textwidth}
         \centering
         \includegraphics[width=\textwidth]{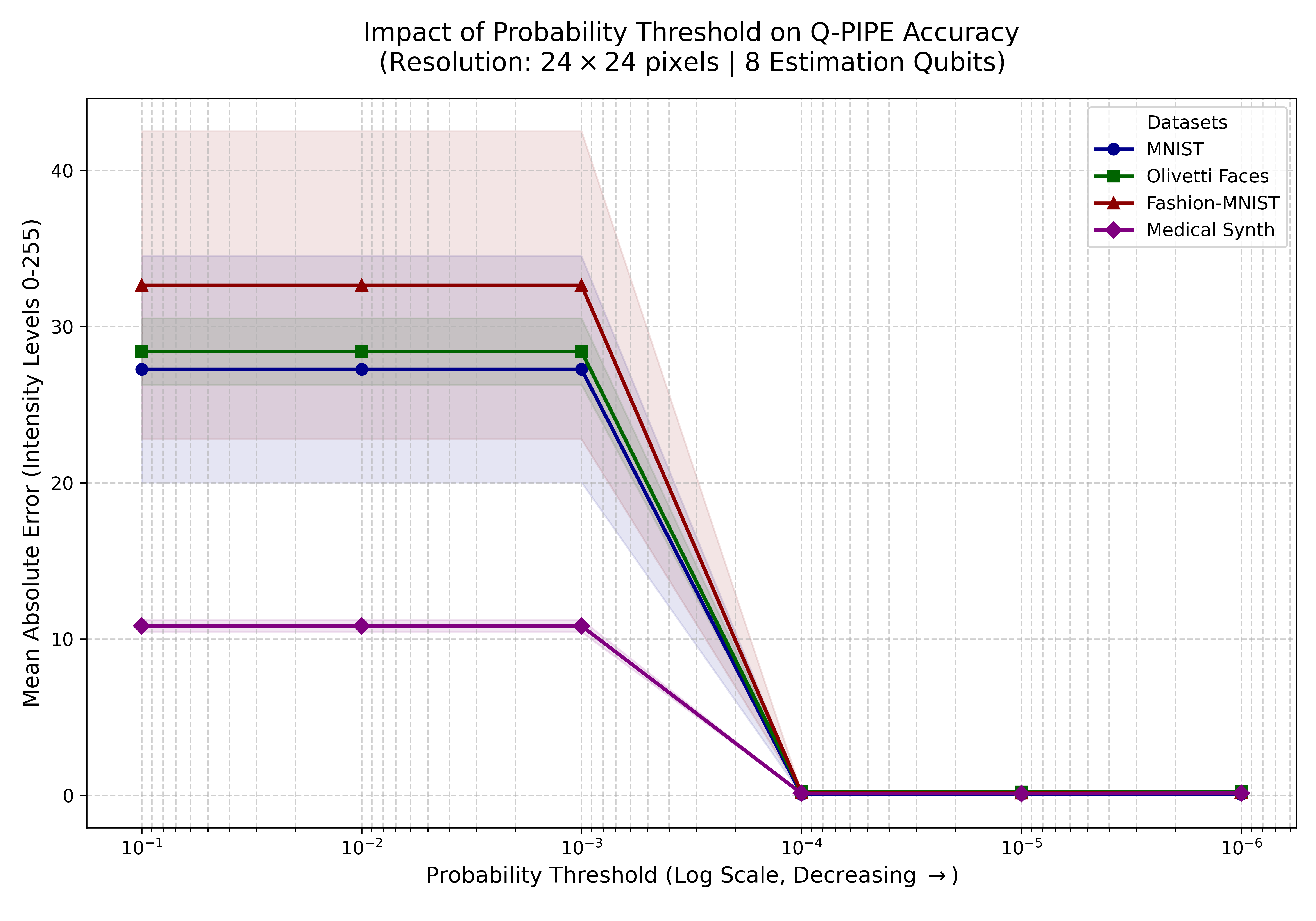}
         \caption{Spatial resolution of $24 \times 24$ pixels.}
         \label{fig:q_pipe_diff_prob_thresholds_24x24}
     \end{subfigure}
    \caption{MAE of the gradient magnitude (lower is better) versus the readout threshold $P_{\mathrm{th}}$ ($10^{-1}$ to $10^{-6}$) at fixed resolution, for MNIST, Olivetti Faces, Fashion-MNIST and the synthetic medical dataset.}
    \label{fig:q_pipe_diff_prob_thresholds}
\end{figure*}

The degradation is a readout effect of the fixed threshold, predicted by the main-peak condition of section~\ref{res:q_pipe_benchmark_prob_threshold}: the $20\times20$ case ($n=9$) gives $2^{n}P_{\mathrm{th}} \approx 0.26 < 4/\pi^2$, whereas the $24\times24$ case ($n=10$) gives $2^{n}P_{\mathrm{th}} \approx 0.51 > 4/\pi^2$, so central QPE peaks begin to be discarded. The threshold must therefore scale as $2^{-n}$.

\subsection{Probability threshold} \label{res:q_pipe_benchmark_prob_threshold}

With the position register in uniform superposition, each address is observed with probability $2^{-n}$, so $P(k,x,y) \le 2^{-n}$. Finite-resolution QPE spreads this probability over neighboring intensity bins according to a Dirichlet kernel \cite{cleve1998quantum, NielsenChuang}, and the side lobes satisfy $P(k,x,y) \ll 2^{-n}$. A fixed threshold therefore removes an $n$-dependent fraction of the signal. Writing $2^{q}\Delta\theta = b + \delta$, with $b$ the nearest integer and $|\delta| \le \tfrac{1}{2}$, the outcome $b+j$ is observed with probability

\begin{equation}
    p_q(j;\delta) = \frac{\sin^2(\pi\delta)}{2^{2q}\sin^2\!\left(\pi(j-\delta)/2^{q}\right)}
    \;\xrightarrow{\;|j| \ll 2^{q}\;}\;
    \frac{\sin^2(\pi\delta)}{\pi^2 (j-\delta)^2},
    \label{eq:qpe_kernel}
\end{equation}

so that $P(b+j,x,y) = 2^{-n}\,p_q(j;\delta)$. Two consequences fix the threshold. First, $p_q(0;\delta) \ge 4/\pi^2 \approx 0.405$ for every $|\delta|\le\tfrac{1}{2}$, so $P_{\mathrm{th}} < 4/(\pi^2 2^{n})$ guarantees that the central peak is never discarded; for large $q$ this condition is also necessary. Second, bounding $\sin^2(\pi\delta)\le 1$ and $(j-\delta)^2 \approx j^2$, the $j$-th side lobe has envelope $1/(\pi^2 j^2)$. Retaining all side lobes up to order $J$ therefore requires

\begin{equation}
    P_{\mathrm{th}} \le \frac{C_D}{2^{n}}, \qquad C_D = \frac{1}{\pi^2 J^2}.
    \label{eq:pth_bound}
\end{equation}

We recommend $J=2$, which retains the two nearest side lobes on each side of the peak and gives $C_D = 1/(4\pi^2) \approx 0.025$. The choice of $J$ is a design parameter rather than a unique optimum, since the exact lobe weight depends on the unknown offset $\delta$. The fixed thresholds used in sections~\ref{res:q_pipe_benchmark} and \ref{res:q_pipe_benchmark_resolution} are not scaled; they satisfy the main-peak condition for $n\le9$ ($5\times10^{-4}$) and $n\le8$ ($10^{-3}$).

We validate \eqref{eq:pth_bound} by sweeping $P_{\mathrm{th}}$ from $10^{-1}$ to $10^{-6}$ at $22\times22$ ($n=9$) and $24\times24$ ($n=10$) pixels (figure~\ref{fig:q_pipe_diff_prob_thresholds}). For $n=9$, the baseline probability is $2^{-9}\approx 1.95\times10^{-3}$. Thresholds of $10^{-1}$ and $10^{-2}$ exceed it and remove the signal (MAE $\approx 13$--$35$). At $10^{-3}$, $2^{n}P_{\mathrm{th}} \approx 0.51 > 4/\pi^2$, so the central peak is discarded whenever $|\delta|$ is large and recovery is only partial. For $P_{\mathrm{th}} \le 10^{-4}$ ($2^{n}P_{\mathrm{th}} \approx 0.05$), at least the first side lobe is retained and the MAE reaches stable, near-optimal values, consistent with \eqref{eq:pth_bound} for $J \approx 1$--$2$. For $n=10$, $P_{\mathrm{th}}=10^{-3}$ gives $2^{n}P_{\mathrm{th}} \approx 1.02$, above the largest possible value of $p_q$, so every outcome is filtered out (MAE $\approx 11$--$33$). At $10^{-4}$, $2^{n}P_{\mathrm{th}} \approx 0.10 \approx 1/\pi^2$, the first-side-lobe boundary, and near-optimal accuracy is recovered.


\subsection{Complexity analysis} \label{sec:complexity}

We compare the resources of Q-PIPE with representative FRQI and NEQR state-preparation circuits \cite{Le2011, Zhang2013, khan2019improved, amankwah2022quantum}. Here $N=2^n$ is the number of pixels (other sizes are zero-padded to the next power of two), $q$ the number of estimation qubits and $N_{\neq0}\le N$ the number of nonzero pixels. All methods use the same $q$-bit intensity precision, which in FRQI sets the rotation angles rather than the number of qubits. We separate two abstraction levels. At the high level, each $n$-controlled phase gate $C^nP$ counts as one operation. At the elementary level (arbitrary single-qubit gates and CNOTs), each $C^nP$ costs $g_{\mathrm{MCP}}(n)$ gates and depth $d_{\mathrm{MCP}}(n)$, which depend on the synthesis strategy, gate set, connectivity and ancilla assumptions \cite{shende2006synthesis, dasilva2022linear}. The Hadamard layer costs $q+n$ gates and the $\mathrm{QFT}^\dagger$ costs $O(q^2)$ \cite{zhou2017quantum}; the dependence on image size comes from the oracle.

\subsubsection{Naive implementation:}

In the naive construction, each address is selected independently for each estimation qubit: $X$ gates map it to the all-ones control state and are then uncomputed. On average half of the address bits are zero, so selection costs $n$ $X$ gates per pixel, $G_X^{\mathrm{naive}} = qnN$, and $G_{\mathrm{MCP}}^{\mathrm{naive}} = qN_{\neq0} \le qN$ phase gates are needed, one per nonzero pixel and estimation qubit. Counting each $C^nP$ as one gate,

\begin{equation}
    G_{\mathrm{high}}^{\mathrm{naive}} = (q+n) + qnN + qN_{\neq 0} + O(q^2) = O(qN\log N).
    \label{eq:naive_total}
\end{equation}

At the elementary level, the phase term is multiplied by its synthesis cost,

\begin{equation}
    G_{\mathrm{elem}}^{\mathrm{naive}} = (q+n) + qnN + qN_{\neq 0}\,g_{\mathrm{MCP}}(n) + O(q^2).
    \label{eq:naive_elementary}
\end{equation}

For exact ancilla-free synthesis of multi-controlled $\mathrm{U}(2)$ gates such as $C^{n}P$, no linear-size construction is known; with current constructions the CNOT count scales as $\Theta(n^{2})$ (Qiskit yields $6$, $44$, $324$ and $1732$ CNOTs for $n=2$, $4$, $8$, $16$), so that $G_{\mathrm{elem}}^{\mathrm{naive}}=O(qN\log^{2}N)$, while linear depth remains achievable \cite{dasilva2022linear}. The selection $X$ gates of one pixel act on distinct qubits and run in parallel, so, with $D_{\mathrm{QFT}}(q)$ the depth of the $\mathrm{QFT}^\dagger$,

\begin{equation}
    D_{\mathrm{naive}} = O\big(qN+qN_{\neq 0}\,d_{\mathrm{MCP}}(n)+D_{\mathrm{QFT}}(q)\big),
    \label{eq:naive_depth}
\end{equation}

which reduces to $O(qN\log N)$ for a linear-depth synthesis, $d_{\mathrm{MCP}}(n)=O(n)$ \cite{dasilva2022linear}, and $N_{\neq 0}=\Theta(N)$.

\subsubsection{Gray-code traversal:}

The Gray-code oracle (figure~\ref{fig:qpipe_gray_oracle}) visits the addresses in binary-reflected Gray-code order, $g_s = s\oplus(s\gg1)$ for $s=0,\ldots,N-1$, and applies $C^nP(2\pi\,2^j\theta_{g_s})$ at each nonzero pixel. For each estimation qubit, it needs $n$ $X$ gates to map $g_0=0$ to the all-ones state, one $X$ gate per transition because consecutive states differ in one bit, and $n-1$ $X$ gates to restore the register from $g_{N-1}=2^{n-1}=10\cdots0$:

\begin{equation}
    G_X^{\mathrm{Gray}} = q\big[\,n + (N-1) + (n-1)\,\big] = q(2n+N-2).
    \label{eq:gray_X_per_qubit}
\end{equation}

\begin{figure*}[t]
     \begin{subfigure}[b]{0.32\textwidth}
         \centering
         \includegraphics[width=\textwidth]{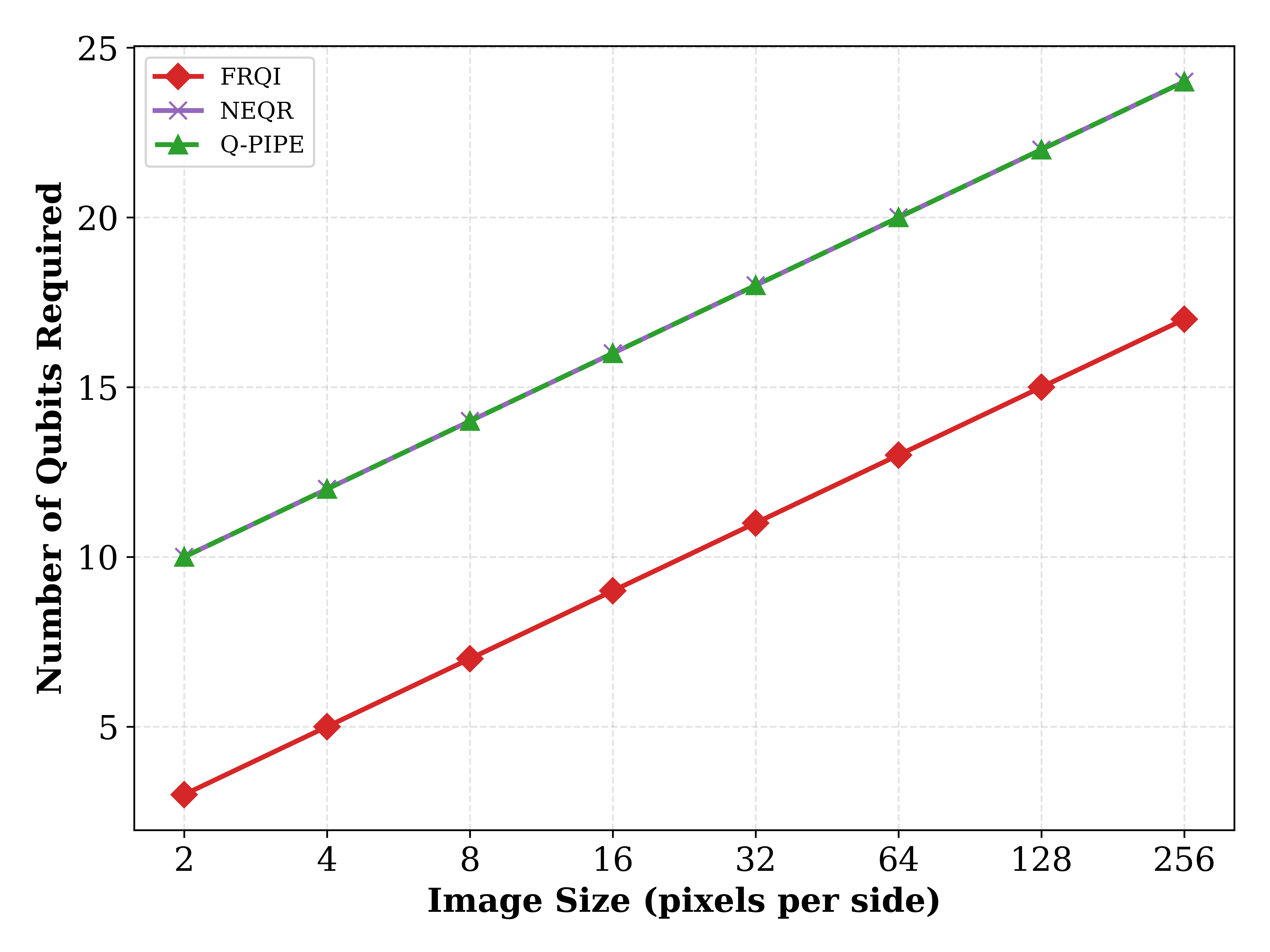}
         \caption{Qubit requirements}
         \label{fig:complexity_qubits}
     \end{subfigure}
     \hfill
     \begin{subfigure}[b]{0.32\textwidth}
         \centering
         \includegraphics[width=\textwidth]{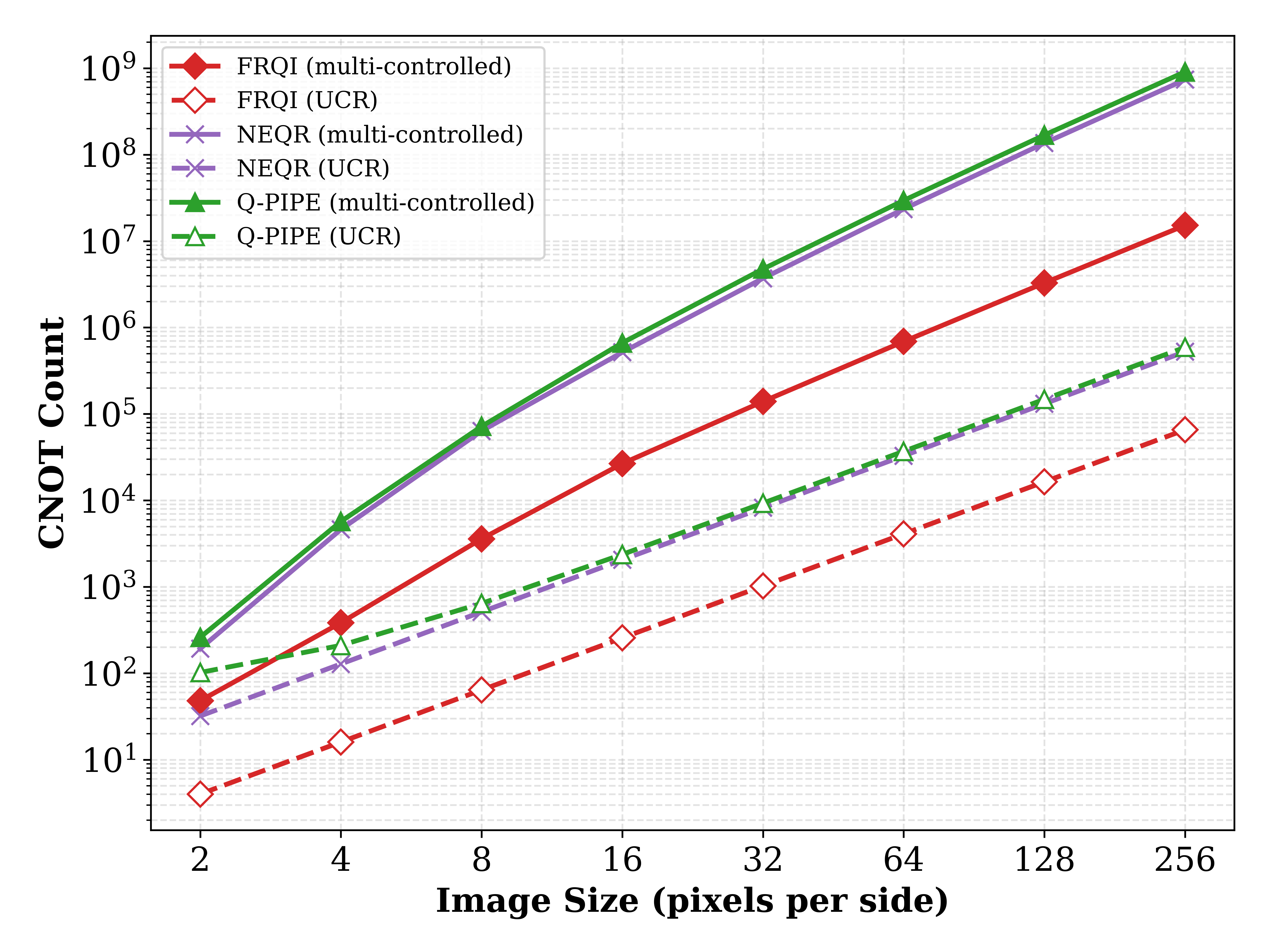}
         \caption{CNOT count}
         \label{fig:complexity_gates}
     \end{subfigure}
     \hfill
     \begin{subfigure}[b]{0.32\textwidth}
         \centering
         \includegraphics[width=\textwidth]{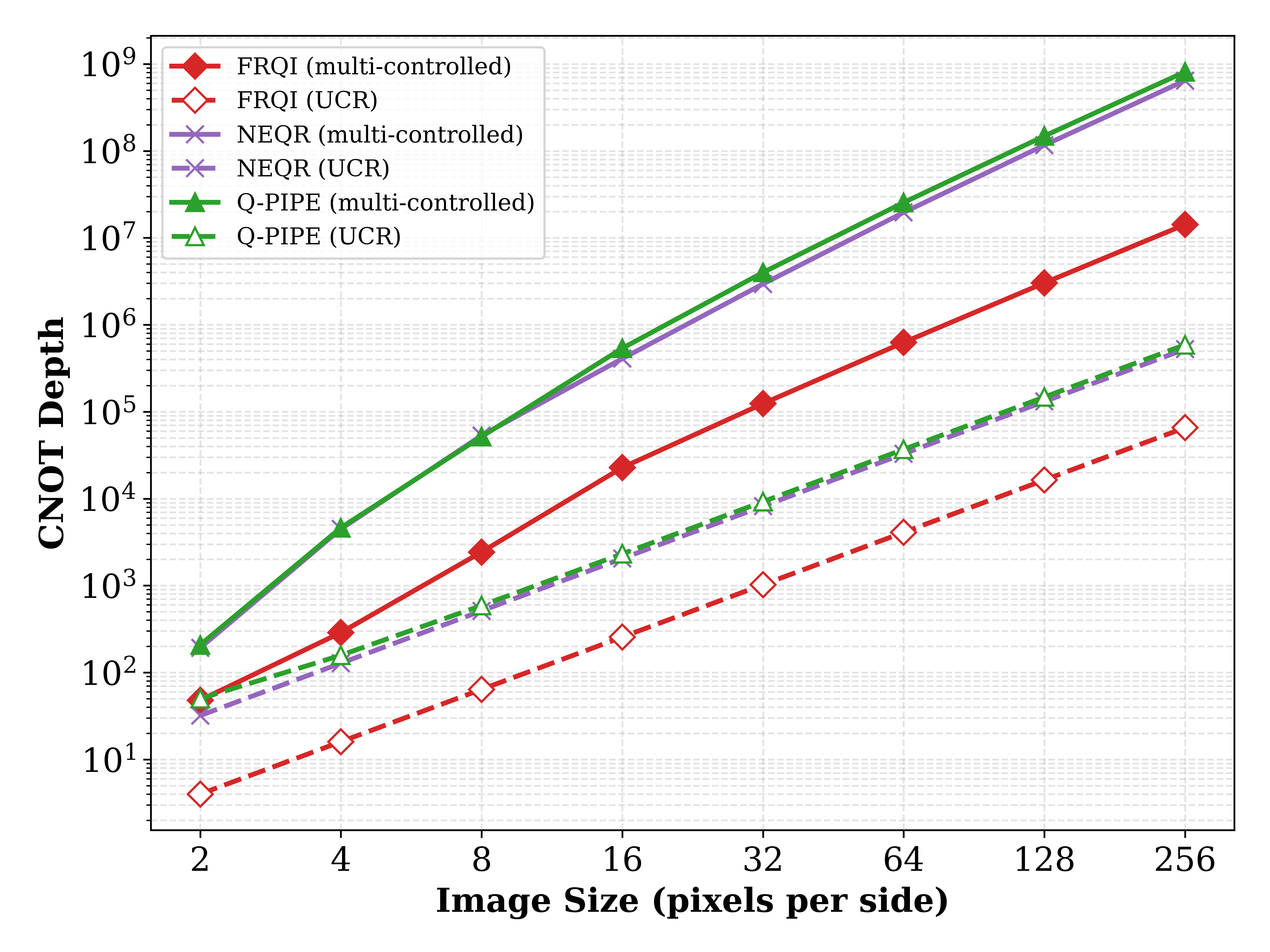}
         \caption{CNOT depth}
         \label{fig:complexity_depth}
     \end{subfigure}
    \caption{Resource scaling for $w\times w$ images ($w=2,\ldots,256$), $q=8$, in one elementary gate set (CNOT and single-qubit gates, no ancillas), worst case (all pixels and intensity bits nonzero). Solid: multi-controlled constructions, one $C^{n}R_y$ (FRQI), $C^{n}X$ (NEQR) or $C^{n}P$ (Q-PIPE, figure~\ref{fig:qpipe_gray_oracle}) per pixel, synthesized with current ancilla-free decompositions (Qiskit~2.5). Dashed: using UCR optimization, \eqref{eq:ucr_total} for Q-PIPE, NEQR and FRQI \cite{mottonen2004quantum, amankwah2022quantum}.}
    \label{fig:complexity_analysis}
\end{figure*}

The number of phase gates is unchanged, $G_{\mathrm{MCP}}^{\mathrm{Gray}} = qN_{\neq 0}$, and no qubits are added: Q-PIPE uses $n+q$ qubits, like NEQR \cite{Zhang2013}, against $n+1$ for FRQI \cite{Le2011} (figure~\ref{fig:complexity_qubits}). The totals are

\begin{equation}
    G_{\mathrm{high}}^{\mathrm{Gray}} = (q+n) + q(2n+N-2) + qN_{\neq 0} + O(q^2) = O(qN),
    \label{eq:gray_total}
\end{equation}

\begin{equation}
    G_{\mathrm{elem}}^{\mathrm{Gray}} = (q+n) + q(2n+N-2) + qN_{\neq 0}\,g_{\mathrm{MCP}}(n) + O(q^2),
    \label{eq:gray_elementary}
\end{equation}

so the $O(qN)$ of \eqref{eq:gray_total} is a high-level count, not an elementary-gate complexity. The Gray code reduces only the selection gates, by the factor

\begin{equation}
    R_X(N) \equiv \frac{G_X^{\mathrm{naive}}}{G_X^{\mathrm{Gray}}} = \frac{nN}{2n+N-2} \;\simeq\; \log_2 N \quad (N\gg n),
    \label{eq:reduction}
\end{equation}

and leaves the $qN_{\neq0}$ phase gates untouched. Transitions have unit depth, but successive pixel steps act on overlapping qubits and are sequential, so

\begin{equation}
    D_{\mathrm{Gray}} = O\big(qN+qN_{\neq 0}\,d_{\mathrm{MCP}}(n)+D_{\mathrm{QFT}}(q)\big),
    \label{eq:gray_depth}
\end{equation}

which is again $O(qN\log N)$ for $d_{\mathrm{MCP}}(n)=O(n)$ and $N_{\neq 0}=\Theta(N)$: the Gray code alone does not reduce the asymptotic depth.

These counts characterize particular circuits, not representations. FRQI preparation drops from $O(N^2)$ \cite{Le2011} to $O(N\log N)$ with $n-2$ ancillas \cite{khan2019improved} and to $O(N)$ with UCR \cite{amankwah2022quantum}; NEQR preparation drops from $O(qN\log N)$ with $n-2$ ancillas \cite{Zhang2013} to $O(qN)$ \cite{amankwah2022quantum}. The same synthesis applies to the Q-PIPE oracle (section~\ref{sec:ucr}).


\subsubsection{Uniformly controlled synthesis of the oracle:} \label{sec:ucr}

The synthesis dependence of \eqref{eq:naive_elementary} and \eqref{eq:gray_elementary} can be removed. Because $U_{\mathrm{img}}$ is diagonal, each controlled power is diagonal on $(e_j,P)$ and factorizes as

\begin{equation}
    \begin{aligned}
        C\text{-}U_{\mathrm{img}}^{2^{j}} &= R_{z}^{(e_j)}\big(\boldsymbol{\varphi}^{(j)}\big)\,\big(I\otimes D_j\big),\\
        \varphi^{(j)}_{x} &= 2\pi\, 2^{j}\theta_x, \qquad D_j=\mathrm{diag}_x\big(e^{i\varphi^{(j)}_{x}/2}\big),
    \end{aligned}
    \label{eq:ucr_factor}
\end{equation}

where $R_{z}^{(e_j)}(\boldsymbol{\varphi}^{(j)})$ applies $R_z(\varphi^{(j)}_x)$ to $e_j$, uniformly controlled by the position register. For QED, $\theta_x$ is replaced by $\Delta\theta_x$, so $U_{\mathrm{img},A}U_{\mathrm{img},B}^{\dagger}$ is a single diagonal. The $q$ diagonals $D_j$ commute and merge into one $n$-qubit diagonal. Each UCR requires $2^{n}$ CNOTs in a Gray-code-ordered parity network \cite{mottonen2004quantum}, the construction behind linear-cost FRQI and NEQR \cite{amankwah2022quantum}, and the merged diagonal requires $2^{n}-2$ CNOTs \cite{bullock2004asymptotically, welch2014efficient}. Without ancillas,

\begin{equation}
    G_{\mathrm{CX}}^{\mathrm{UCR}} = (q+1)N-2+G_{\mathrm{CX}}^{\mathrm{QFT}^{\dagger}}(q) = O(qN), \qquad D_{\mathrm{CX}}^{\mathrm{UCR}} = O(qN).
    \label{eq:ucr_total}
\end{equation}

This circuit implements exactly the unitary of the $C^nP$ oracle (figure~\ref{fig:qpipe_gray_oracle}), the results of section~\ref{sec: Results and Simulations} are unchanged; the Gray code is retained, but now orders the parity network instead of the $X$ gates.

Linear preparation cost is therefore a property of the uniformly controlled synthesis of block-diagonal (FRQI, NEQR) and diagonal (Q-PIPE) oracles, not of a particular representation. With both syntheses in one gate set (figure~\ref{fig:complexity_analysis}), every representation drops to $\Theta(N)$ CNOTs per intensity register: $N$ (FRQI), $qN$ (NEQR) and $(q+1)N$ (Q-PIPE). FRQI remains the cheapest per execution, but its amplitude readout requires $O(4^{q})$ shots per pixel for $q$-bit precision, against $O(1)$ for the basis-state outputs of NEQR and Q-PIPE \cite{amankwah2022quantum, cleve1998quantum}. Q-PIPE and NEQR therefore have practically the same preparation cost, and the difference between them is functional (section~\ref{sec:discussion}).

\subsubsection{Readout complexity} \label{sec:readout}

The counts above describe a single execution. Recovering the full classical image requires repeated measurement: each shot samples a position with probability $1/N$, so estimating every pixel to additive precision $\varepsilon$ with failure probability at most $\delta$ (a Hoeffding bound with a union bound over pixels) requires, up to constant factors,

\begin{equation}
    M = O\!\left(\frac{N}{\varepsilon^2}\log\frac{N}{\delta}\right)
    \label{eq:readout_shots}
\end{equation}

repetitions, or $M=O(N/\varepsilon^2)$ under an average per-pixel criterion. Any representation in which each shot reveals one sampled position needs $\Omega(N)$ shots, so readout offsets the logarithmic qubit compression whenever the whole image must be reconstructed classically. This cost is avoided when the Q-PIPE state is passed coherently to a subsequent quantum routine, such as a variational layer, in which only a task-specific observable is measured.


\section{Discussion and outlook} \label{sec:discussion}

With uniformly controlled synthesis, Q-PIPE matches the $O(qN)$ preparation cost of optimized NEQR (section~\ref{sec:ucr}); the compiled depth of the oracle, not the phase-encoding principle, therefore sets the present NISQ boundary (appendix~\ref{app:noise}).

The distinction from NEQR is functional. For inputs commensurate with the estimation register, Q-PIPE produces the same position--intensity state as NEQR, so the two are interchangeable as data loaders; they differ in the operations available in place. In NEQR, intensities reside in computational-basis states, and arithmetic between images, such as the directional differences underlying edge detection, requires a quantum adder on the intensity register: ripple-carry designs need ancillas \cite{vedral1996quantum}, and Fourier-space adders \cite{draper2000addition, ruizperez2017quantum} need a forward and $QFT^{\dagger}$ around the addition. In Q-PIPE, intensities are carried as phases, loading and Fourier-space addition are fused into a single pass that needs only the final $\mathrm{QFT}^{\dagger}$; sequential oracle application adds or subtracts them by construction (equations \ref{eq:utotal_basis}--\ref{eq:qed_qpipe}), and the $\mathrm{QFT}^{\dagger}$ reads out the result without an arithmetic circuit. Existing phase- and frequency-domain representations \cite{lisnichenko2023quantum, Wereszczynski2020, grigoryan2025image, grigoryan2020new, alwan2025multilayered} also store data in phases or spectral coefficients, but as a static medium that must be transformed back before pixel-wise processing; in Q-PIPE, the phases are the substrate of the computation.

Two classical corrections make the readout robust at finite register size. The half-spectrum mapping confines phase differences to $[-\pi,\pi]$ and, with the decoding rule \eqref{eq:decoding}, recovers both signs of the gradient \cite{spagnolini19932, zhu2022anti, NielsenChuang}. Spectral leakage is handled by probability-weighted reconstruction rather than single maximum-likelihood outcomes \cite{li2024iterative, harris1978use}, with a threshold $P_{\mathrm{th}} \le C_D/2^{n}$ derived from the QPE kernel \eqref{eq:pth_bound} and validated by threshold sweeps (section~\ref{res:q_pipe_benchmark_prob_threshold}). In ideal simulation, Q-PIPE thus gives zero MAE for inputs commensurate with the estimation register and MAE $\lesssim 0.35$ up to $20\times20$ pixels for continuous data, and the degradation at $24\times24$ is removed by lowering the threshold.

Under three IBM device noise models (appendix~\ref{app:noise}), the ideal pipeline reproduces the classical gradient (MAE $\approx 0.22$), whereas the $C^{n}P$ implementation is driven toward the maximally mixed state ($F\approx0.10$, MAE $\approx72$) by the depth of its multi-controlled-phase synthesis. The equivalent UCR implementation reduces the transpiled depth by a factor of $53$--$59 \times$, raises the classical fidelity to $0.56$--$0.70$ and preserves the edge structure (Pearson $r=0.99$ with the classical gradient), leaving a residual MAE of $\approx18$--$28$.

\section{Conclusions} \label{sec:conclusions}

We have introduced Q-PIPE, a phase-encoding framework that loads continuous data as eigenphases of a diagonal oracle, coherently maps them to computational-basis states via QPE, and executes arithmetic operations directly in the phase domain. By exploiting phase kickback followed by the $QFT^{\dagger}$, equivalent to Fourier-space addition; Q-PIPE is, to our knowledge, the first data-loading scheme to unify coherent phase-to-basis readout with native arithmetic between datasets without auxiliary arithmetic circuits.

The framework is practical in terms of quantum resources. When synthesized with UCRs, the oracle requires $O(qN)$ CNOT gates without ancillas, matching the asymptotic scaling of optimized NEQR preparation, while basis-state readout requires only $O(1)$ shots per pixel at $q$-bit precision. Applied to QED, Q-PIPE computes directional gradients exactly for inputs commensurate with the estimation register, and with errors below one intensity level for arbitrary continuous data in ideal simulations. Furthermore, an analytical threshold rule derived from the phase-estimation kernel systematically characterizes and eliminates resolution-dependent readout errors. Under realistic device noise models, this rotation-based synthesis robustly preserves edge structures where standard multi-controlled gate constructions fail.

Future work will focus on reducing circuit depth through approximate $\mathrm{QFT}^\dagger$ implementations, truncated estimation registers, and error mitigation. Additionally, Q-PIPE can serve as a coherent data-embedding layer for variational quantum algorithms and quantum kernel methods, where quantum states are consumed coherently, thereby bypassing the classical readout overhead discussed in section~\ref{sec:readout}.

\section*{Acknowledgments}
S.-H. Dong started this work in China. The authors declare no competing interests.






\section*{Data}
The code and data that support the findings of this study are openly available at \url{https://github.com/BrianSarmina/Papers/tree/main/Q-PIPE}. 
A PennyLane community demo of Q-PIPE is available at \url{https://pennylane.ai/demos_community}.

\appendix

\renewcommand{\thefigure}{\thesection\arabic{figure}}
\renewcommand{\thetable}{\thesection\arabic{table}}
\setcounter{figure}{0}

\section{Quantum edge detection on MNIST} \label{app:QED_benchmarks}

\begin{figure*}[t!]
    \centering
    \includegraphics[width=\textwidth]{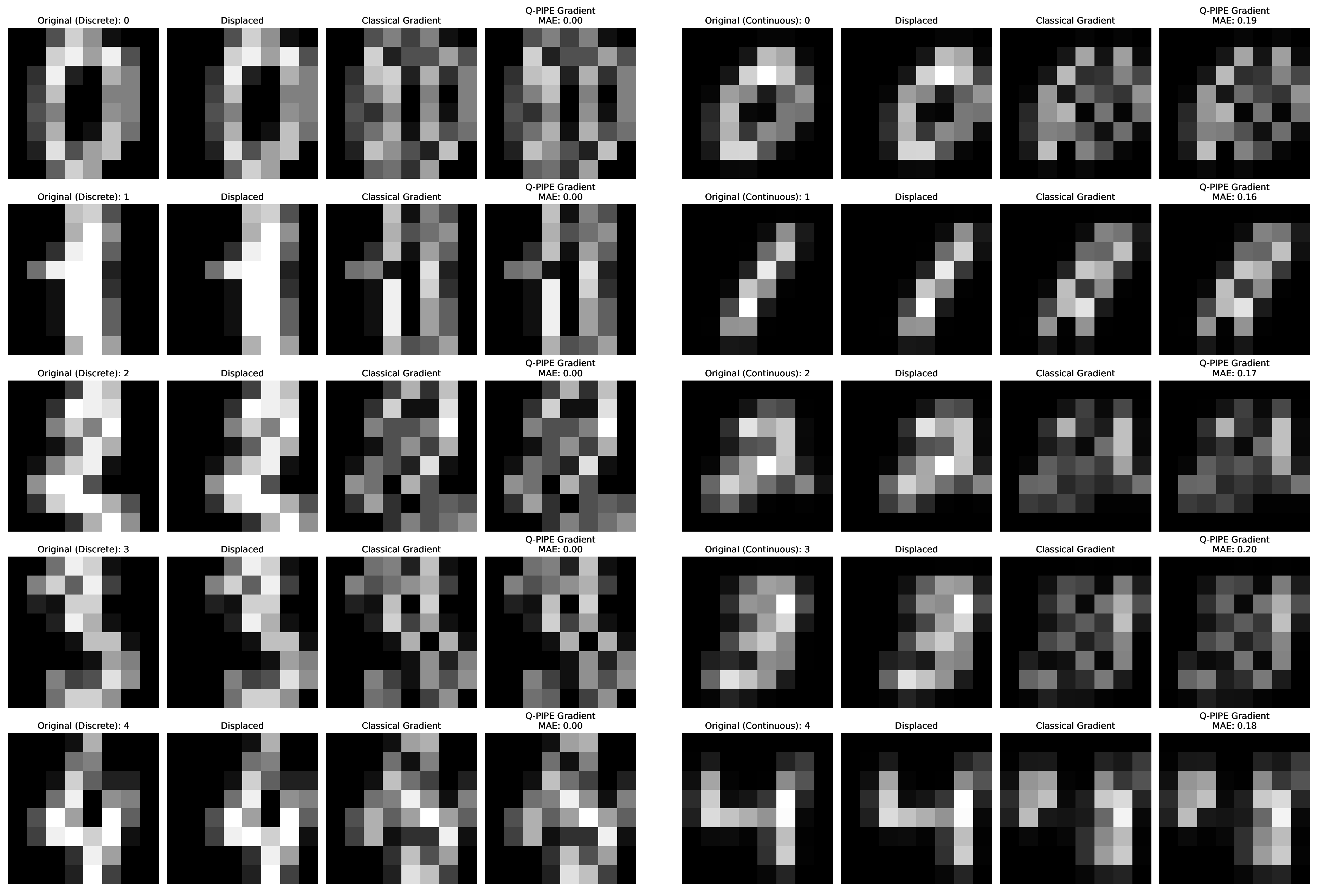}
    \caption{Horizontal-gradient QED on MNIST at $8\times8$ resolution. Left four columns: \textit{load\_digits} (discrete); right four columns: \textit{fetch\_openml} (continuous). In each block, from left to right: original image, horizontally shifted image, classical gradient, and Q-PIPE gradient with its MAE.}
    \label{fig:s1}
\end{figure*}

\begin{figure*}[t!]
    \centering
    \includegraphics[width=\textwidth]{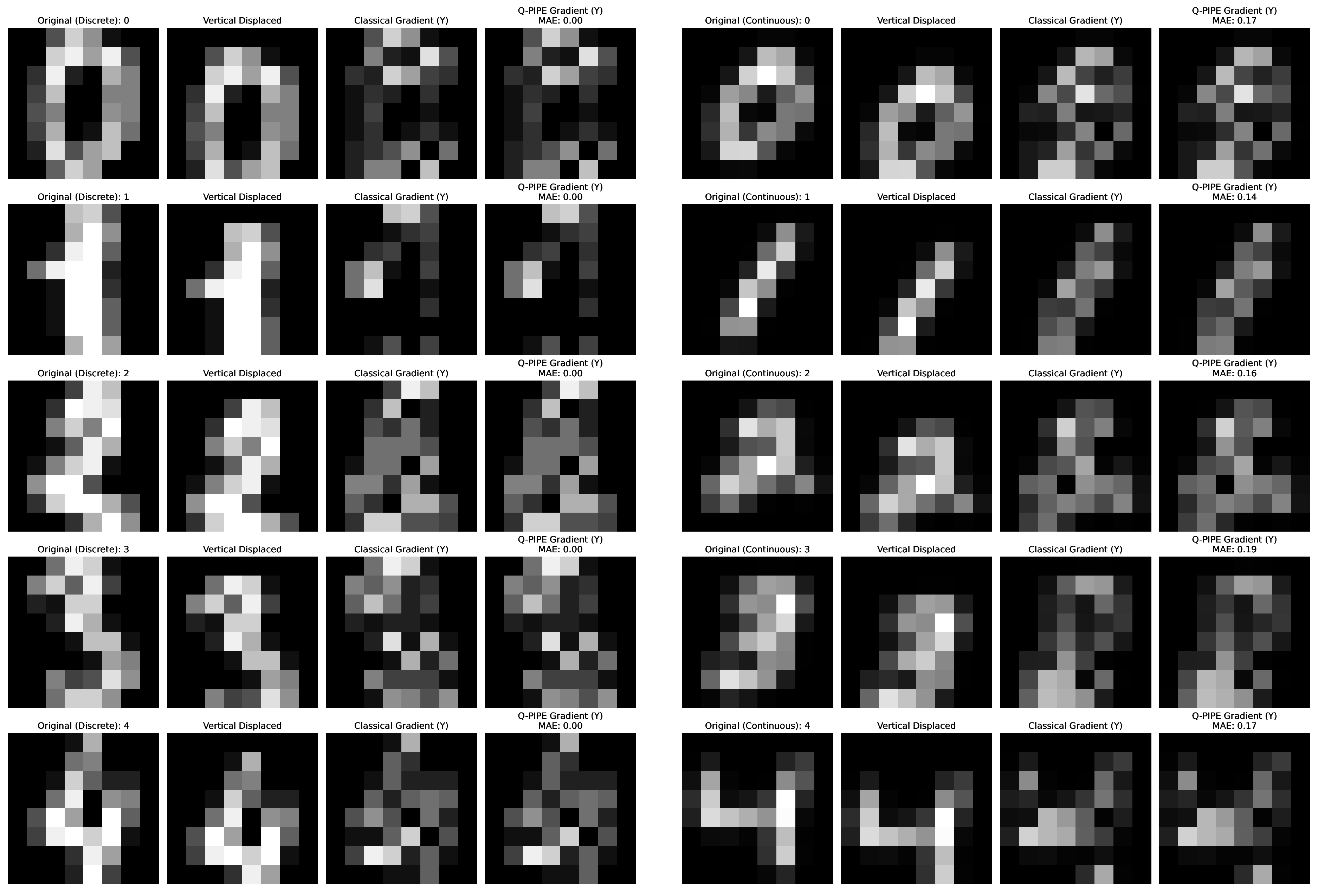}
    \caption{Vertical-gradient QED on MNIST at $8\times8$ resolution. Panel layout as in figure~\ref{fig:s1}, with the image shifted vertically.}
    \label{fig:s2}
\end{figure*}

\begin{figure*}[t!]
    \centering
    \includegraphics[width=\textwidth]{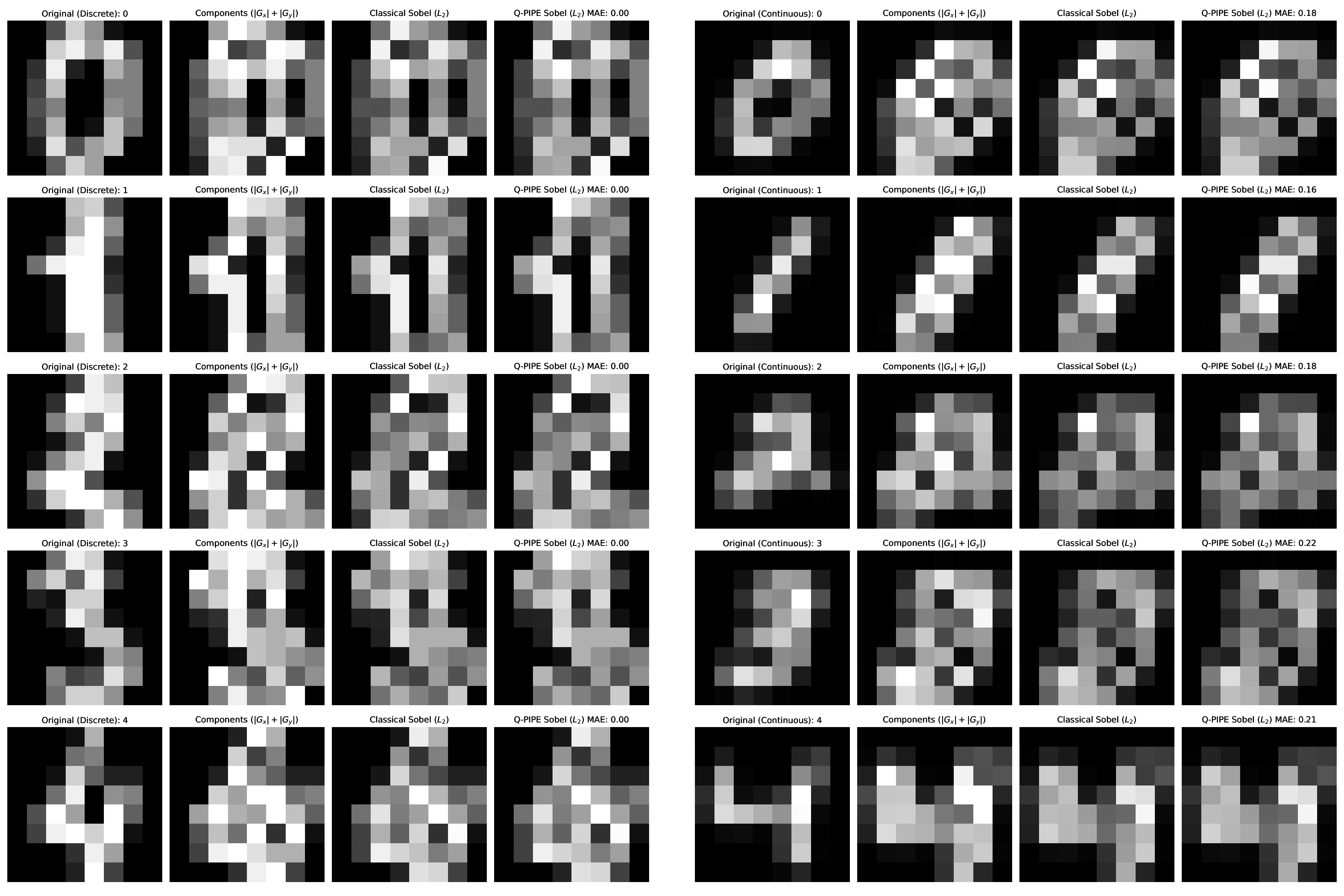}
    \caption{Gradient magnitude on MNIST at $8\times8$ resolution. Left four columns: \textit{load\_digits}; right four columns: \textit{fetch\_openml}. In each block, from left to right: original image, sum of the directional components $|G_x|+|G_y|$, classical gradient magnitude, and Q-PIPE gradient magnitude \eqref{eq:sobel} with its MAE (labelled ``Sobel'' in the panels).}
    \label{fig:s3}
\end{figure*}

This appendix shows representative QED reconstructions for the two MNIST variants provided by \textit{scikit-learn}: the native $8\times8$ \textit{load\_digits} set and the $28\times28$ \textit{fetch\_openml} set, downsampled to $8\times8$. Figures~\ref{fig:s1}, \ref{fig:s2} and \ref{fig:s3} show the horizontal gradient, the vertical gradient and the gradient magnitude, respectively.

For \textit{load\_digits}, the MAE is exactly zero in all three cases. Its intensities take only the 17 levels $I = m\,I_{\max}/16$ ($m=0,\ldots,16$), so every difference gives $\Delta\theta = m'/32$ with integer $m'$; $2^{q}\Delta\theta$ is therefore an integer for all $q\ge5$, and QPE returns each gradient exactly as a computational-basis state.

For \textit{fetch\_openml}, the downsampled intensities are generally not commensurate with the estimation register, QPE spreads each value over neighboring basis states. With the probability-weighted reconstruction of section~\ref{sec: Results and Simulations}, the MAE remains below one intensity level: $0.16$--$0.20$ for the horizontal gradient, $0.14$--$0.19$ for the vertical gradient and $0.16$--$0.22$ for the gradient magnitude.

\setcounter{figure}{0}\setcounter{table}{0}

\section{Performance under device noise models} \label{app:noise}

The main-text benchmarks use ideal state-vector simulation. Here we run the encoding and QED pipeline under device noise models, to quantify the gap between ideal and currently attainable accuracy, and to identify its dominant mechanism.

We evaluate horizontal QED on five continuous MNIST samples (\textit{fetch\_openml}, $28\times28$ downsampled to $4\times4$), with the half-spectrum mapping of section~\ref{sec: Results and Simulations}, an 8-qubit estimation register and a 4-qubit position register (12 qubits), 4096 shots and transpilation at optimization level~3. Two constructions of the same unitary are compared: the $C^{n}P$ Gray-code oracle (figure~\ref{fig:qpipe_gray_oracle}) and the UCR synthesis (section~\ref{sec:ucr}). Noise is injected through three fake backends, FakeKingston, FakeMarrakesh and FakeBerlin, whose calibration data (gate and readout errors, $T_1/T_2$ and heavy-hex coupling maps) were synchronized to live IBM Quantum properties at the time of each run. For each instance we compute (i) the classical fidelity $F=\sum_{b}\sqrt{P_{\mathrm{ideal}}(b)\,P_{\mathrm{noise}}(b)}$ (Bhattacharyya coefficient) between the sampled ideal and noisy output distributions, (ii) the MAE of the noisy horizontal gradient against the classical finite difference, and (iii) the transpiled circuit depth.

\begin{table}[t]
\centering
\caption{Noisy-simulation benchmarks for horizontal QED ($4\times4$ MNIST, 8 estimation and 4 position qubits, 4096 shots, $P_{\mathrm{th}}=5\times10^{-4}$) with the $C^{n}P$ and UCR oracles, averaged over five instances ($\pm$ one standard deviation). $F$: classical (Bhattacharyya) fidelity between ideal and noisy output distributions; $H$: Shannon entropy of the noisy distribution (uniform ceiling 12 bits); Mass${}>P_{\mathrm{th}}$: probability mass above the reconstruction threshold; TVD: total-variation distance from the ideal distribution; Depth: transpiled depth (optimization level 3). Ideal depths are elementary-gate depths without coupling constraints ($C^{n}P$ / UCR); both oracles produce the same ideal distribution.}
\label{tab:noise_summary}
\small
\begin{tabular}{llcccccc}
\toprule
Backend & Oracle & $F$ & MAE & Depth & $H$ (bits) & Mass${}>P_{\mathrm{th}}$ & TVD \\
\midrule
Ideal         & both     & ---               & $0.22\pm0.03$ & $18869$ / $287$ & $5.2$  & $0.965$ & ---    \\
\midrule
FakeKingston  & $C^{n}P$ & $0.100\pm0.010$   & $72.0\pm8.5$  & $53938$ & $11.2$ & $0.263$ & $0.96$ \\
FakeMarrakesh & $C^{n}P$ & $0.103\pm0.008$   & $72.0\pm7.0$  & $53551$ & $11.2$ & $0.262$ & $0.96$ \\
FakeBerlin    & $C^{n}P$ & $0.094\pm0.011$   & $71.3\pm8.7$  & $50494$ & $11.2$ & $0.263$ & $0.96$ \\
\midrule
FakeKingston  & UCR      & $0.646\pm0.008$   & $22.1\pm3.6$  & $953$   & $8.9$  & $0.685$ & $0.58$ \\
FakeMarrakesh & UCR      & $0.557\pm0.014$   & $27.6\pm5.3$  & $1010$  & $9.5$  & $0.625$ & $0.68$ \\
FakeBerlin    & UCR      & $0.696\pm0.016$   & $17.8\pm2.2$  & $854$   & $8.6$  & $0.736$ & $0.52$ \\
\bottomrule
\end{tabular}
\end{table}

\begin{figure*}[t]
\centering
\includegraphics[width=\textwidth]{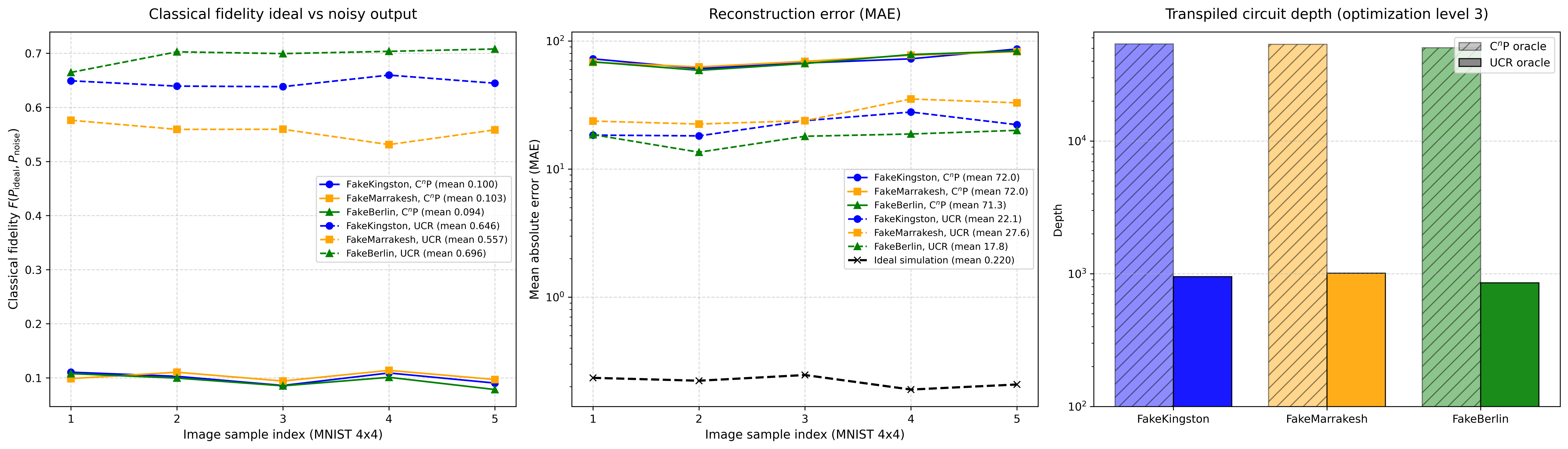}
\caption{Per-instance noisy-simulation metrics for horizontal QED on five $4\times4$ MNIST samples, for the $C^{n}P$ oracle (solid lines, hatched bars) and the equivalent UCR oracle (dashed lines, solid bars) on three synchronized backends. \textbf{Left:} classical (Bhattacharyya) fidelity $F(P_{\mathrm{ideal}},P_{\mathrm{noise}})$. \textbf{Centre:} reconstruction MAE against the classical finite difference (logarithmic scale); the black dashed line is the ideal simulation. \textbf{Right:} mean transpiled depth (logarithmic scale). The UCR synthesis reduces the depth by a factor of $53$--$59$ and raises the fidelity from $F\approx0.10$ to $0.56$--$0.70$.}
\label{fig:noise_metrics}
\end{figure*}

\begin{figure*}[t]
\centering
\includegraphics[width=\textwidth]{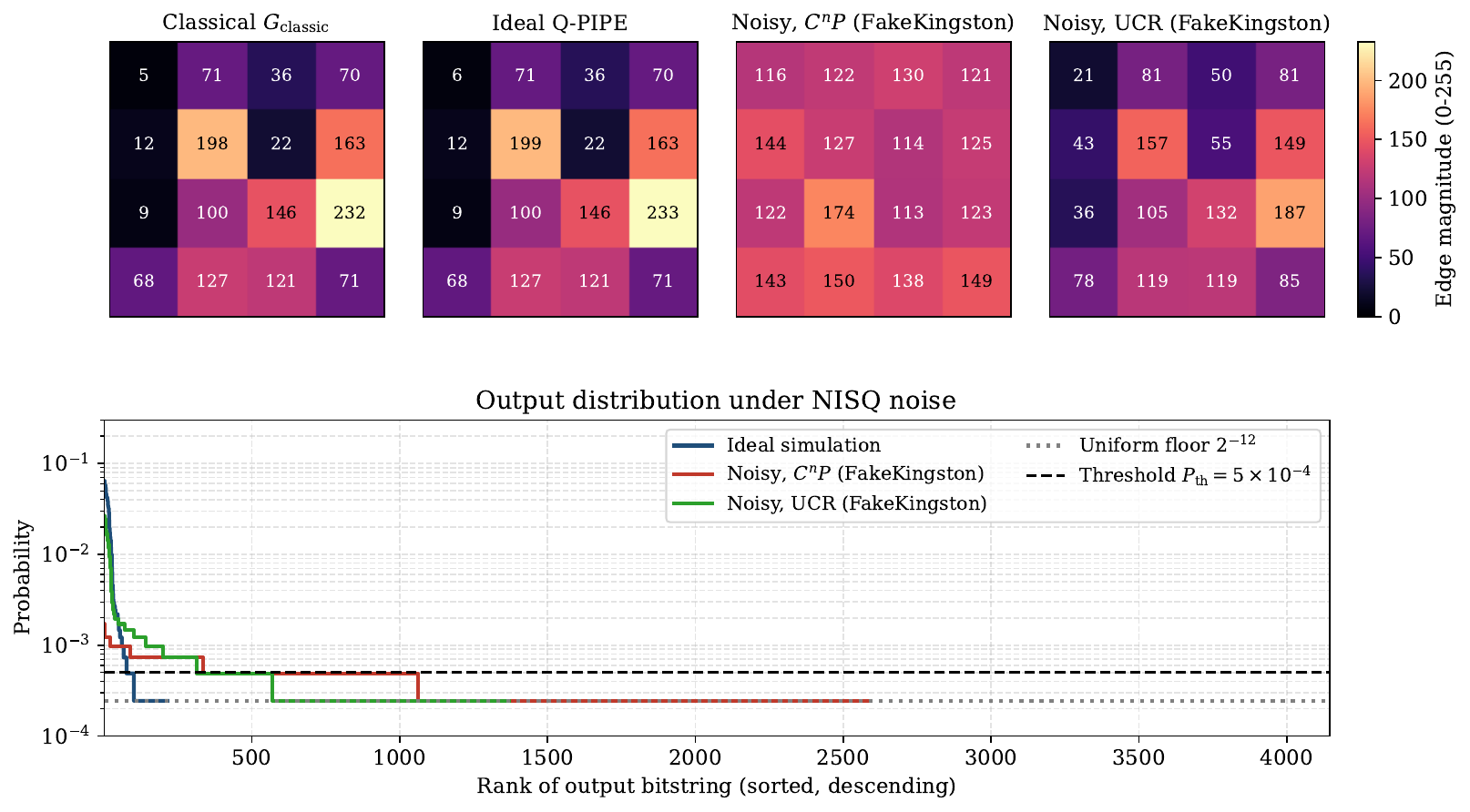}
\caption{Mechanism of noise-induced degradation for both oracle constructions (representative instance, FakeKingston). \textbf{Top:} the classical gradient (left) is reproduced to within one intensity level by the ideal Q-PIPE reconstruction (MAE $\approx0.2$). Under noise, the $C^{n}P$ reconstruction collapses to a nearly flat field near the depolarized mean ($\approx127.5$; MAE $\approx72$, Pearson $r\approx0$), erasing the edge structure, whereas the equivalent UCR reconstruction preserves it (MAE $\approx18$, $r=0.99$) with reduced contrast. \textbf{Bottom:} rank-sorted output probabilities. The $C^{n}P$ distribution flattens toward the uniform floor $2^{-12}$ (dotted; for 4096 shots this also equals a single count), leaving only $\approx27\%$ of the probability mass above the reconstruction threshold $P_{\mathrm{th}}$ (dashed), while the UCR distribution retains its leading peaks and $\approx68\%$ of the mass.}
\label{fig:noise_mechanism}
\end{figure*}

\textbf{Multi-controlled oracle.} The ideal pipeline reproduces the classical gradient (MAE $\approx0.22$, shot-noise limited; table~\ref{tab:noise_summary}). Under noise, the $C^{n}P$ circuits collapse on all three backends to $F\approx0.10$ and MAE $\approx71$--$72$ on the $0$--$255$ scale. The backends are indistinguishable (their means lie within one standard deviation), so the degradation is set by circuit depth rather than by any single calibration (figure~\ref{fig:noise_metrics}). The elementary ideal circuit already has depth $\approx1.9\times10^{4}$ ($\approx9.9\times10^{3}$ CNOTs), because each $C^{n}P$ requires $\Theta(n^{2})$ CNOTs in ancilla-free synthesis (section~\ref{sec:complexity}), and routing onto the heavy-hex connectivity raises it to $\approx5.0$--$5.4\times10^{4}$. The output approaches the maximally mixed state: the entropy rises from $5.2$ to $11.2$~bits, the number of distinct bitstrings from $\approx190$ to $\approx2590$ (of 4096 shots), and the TVD reaches $0.96$. The probability mass above $P_{\mathrm{th}}$ falls from $96.5\%$ to $26.3\%$; the rest lies in a sub-threshold floor near $2^{-12}$ (figure~\ref{fig:noise_mechanism}, bottom). This mechanism differs from the spectral leakage of section \ref{res:q_pipe_benchmark_prob_threshold}: there, probability spreads to \emph{adjacent} intensity bins, whereas here it spreads \emph{globally} toward a depolarized floor. Both, however, act through the same threshold-based readout.

\textbf{Uniformly controlled oracle.} On the same five images, the UCR circuit contains 198 CNOTs (depth 287) instead of $\approx9.9\times10^{3}$ (depth $\approx1.9\times10^{4}$); merging the two oracles into one diagonal accounts for a factor $\approx1.7$ of this reduction and the uniformly controlled synthesis for a factor $\approx29$. After routing, the circuits contain $3.5$--$4.8\times10^{2}$ two-qubit gates instead of $\approx2.1\times10^{4}$, and the transpiled depth falls to $854$--$1010$, a factor $53$--$59$ lower. The fidelity rises to $F=0.56$--$0.70$, the entropy falls to $8.6$--$9.5$~bits, the number of distinct bitstrings to $\approx1180$--$1570$, and the TVD to $0.52$--$0.68$; the mass above $P_{\mathrm{th}}$ rises to $0.63$--$0.74$, and the MAE decreases to $17.8$--$27.6$. The noisy UCR edge maps correlate with the classical gradient with Pearson $r=0.99$ (range $0.97$--$1.00$ over instances and backends), against $r\approx-0.07$ for $C^{n}P$. The residual error is dominated by a systematic contrast compression toward the depolarized mean (linear fit $G_{\mathrm{noisy}}\approx0.64\,G_{\mathrm{classic}}+37$), as expected when a partially depolarized background enters the probability-weighted readout, rather than by a loss of spatial information. Figure \ref{fig:noise_mechanism} illustrates this for a representative instance: the UCR edge map reproduces the spatial pattern of the classical gradient with compressed contrast, and its output distribution retains the leading QPE peaks above $P_{\mathrm{th}}$ instead of flattening toward the uniform floor. The backends are now distinguishable ($F$: FakeBerlin $>$ FakeKingston $>$ FakeMarrakesh), so at this depth device calibration, not circuit depth alone, sets the residual error.

Replacing the multi-controlled synthesis with UCRs, without changing the encoded state, thus turns a nearly depolarized output into an edge map that preserves the gradient structure. The remaining gap, a contrast compression of $\approx36\%$ and an MAE of $\approx18$--$28$, is a concrete target for further depth reduction (approximate or banded $\mathrm{QFT}^{\dagger}$ \cite{barenco1996approximate}, smaller estimation registers traded against intensity precision, as in section~\ref{res:q_pipe_benchmark}, and improved qubit placement) and for error mitigation, such as measurement-error mitigation or zero-noise extrapolation \cite{temme2017error}.

\bibliographystyle{unsrt} 
\bibliography{qpipe_njp_references}

@article{Yan2016,
  author   = {Yan, Fei and Iliyasu, Abdullah M. and Venegas-Andraca, Salvador E.},
  title    = {A survey of quantum image representations},
  journal  = {Quantum Information Processing},
  volume   = {15},
  pages    = {1--35},
  year     = {2016},
  DOI      = {10.1007/s11128-015-1195-6}
}

@article{khan2024beyond,
  author   = {Khan, Mansoor A. and Aman, Muhammad N. and Sikdar, Biplab},
  title    = {Beyond Bits: A Review of Quantum Embedding Techniques for Efficient Information Processing},
  journal  = {IEEE Access},
  volume   = {12},
  pages    = {46118--46137},
  year     = {2024},
  DOI      = {10.1109/access.2024.3382150}
}

@article{FAROOQ2025100763,
  author   = {Farooq, Umar and Singh, Parvinder and Kumar, Atul},
  title    = {A systematic review of quantum image processing: Representation, applications and future perspectives},
  journal  = {Computer Science Review},
  volume   = {57},
  pages    = {100763},
  year     = {2025},
  DOI      = {10.1016/j.cosrev.2025.100763}
}

@article{lisnichenko2023quantum,
  author   = {Lisnichenko, Marina and Protasov, Stanislav},
  title    = {Quantum image representation: a review},
  journal  = {Quantum Machine Intelligence},
  volume   = {5},
  pages    = {2},
  year     = {2023},
  DOI      = {10.1007/s42484-022-00089-7}
}

@article{larose2020robust,
  author   = {LaRose, Ryan and Coyle, Brian},
  title    = {Robust data encodings for quantum classifiers},
  journal  = {Physical Review A},
  volume   = {102},
  pages    = {032420},
  year     = {2020},
  DOI      = {10.1103/physreva.102.032420}
}

@article{weigold2021encoding,
  author   = {Weigold, Manuela and Barzen, Johanna and Leymann, Frank and Salm, Marie},
  title    = {Encoding patterns for quantum algorithms},
  journal  = {IET Quantum Communication},
  volume   = {2},
  pages    = {141--152},
  year     = {2021},
  DOI      = {10.1049/qtc2.12032}
}

@article{ZHANG2026114273,
  author   = {Zhang, Jiale and Che, Xilong and Peng, Shun and Chen, Geng and Ma, Quangong and Fan, Bincheng and Hu, Juncheng},
  title    = {Design and implementation of run-length encoding on quantum computers for resource-efficient data representation},
  journal  = {Applied Soft Computing},
  volume   = {187},
  pages    = {114273},
  year     = {2026},
  DOI      = {10.1016/j.asoc.2025.114273}
}

@inproceedings{weigold2020data,
  author    = {Weigold, Manuela and Barzen, Johanna and Leymann, Frank and Salm, Marie},
  title     = {Data encoding patterns for quantum computing},
  booktitle = {Proceedings of the 27th Conference on Pattern Languages of Programs (PLoP '20)},
  pages     = {1--11},
  year      = {2020}
}

@article{gonzalez2024efficient,
  author   = {Gonzalez-Conde, Javier and Watts, Thomas W. and Rodriguez-Grasa, Pablo and Sanz, Mikel},
  title    = {Efficient quantum amplitude encoding of polynomial functions},
  journal  = {Quantum},
  volume   = {8},
  pages    = {1297},
  year     = {2024},
  DOI      = {10.22331/q-2024-03-21-1297}
}

@article{di2021improving,
  author   = {Di Matteo, Olivia and McCoy, Anna and Gysbers, Peter and Miyagi, Takayuki and Woloshyn, R. M. and Navrátil, Petr},
  title    = {Improving {Hamiltonian} encodings with the {Gray} code},
  journal  = {Physical Review A},
  volume   = {103},
  pages    = {042405},
  year     = {2021},
  DOI      = {10.1103/physreva.103.042405}
}

@article{rath2024quantum,
  author   = {Rath, Minati and Date, Hema},
  title    = {Quantum data encoding: a comparative analysis of classical-to-quantum mapping techniques and their impact on machine learning accuracy},
  journal  = {EPJ Quantum Technology},
  volume   = {11},
  pages    = {72},
  year     = {2024},
  DOI      = {10.1140/epjqt/s40507-024-00285-3}
}

@article{ranga2024quantum,
  author   = {Ranga, Deepak and Rana, Aryan and Prajapat, Sunil and Kumar, Pankaj and Kumar, Kranti and Vasilakos, Athanasios V.},
  title    = {Quantum Machine Learning: Exploring the Role of Data Encoding Techniques, Challenges, and Future Directions},
  journal  = {Mathematics},
  volume   = {12},
  pages    = {3318},
  year     = {2024},
  DOI      = {10.3390/math12213318}
}

@misc{munikote2024comparing,
  author        = {Munikote, Nidhi},
  title         = {Comparing Quantum Encoding Techniques},
  year          = {2024},
  eprint        = {2410.09121},
  archivePrefix = {arXiv}
}

@article{Le2011,
  author   = {Le, Phuc Q. and Dong, Fangyan and Hirota, Kaoru},
  title    = {A flexible representation of quantum images for polynomial preparation, image compression, and processing operations},
  journal  = {Quantum Information Processing},
  volume   = {10},
  pages    = {63--84},
  year     = {2011},
  DOI      = {10.1007/s11128-010-0177-y}
}

@misc{ko2025quantum,
  author        = {Ko, Taehee and Lee, Inho and Yu, Hyeong Won},
  title         = {Quantum medical image encoding and compression using {Fourier}-based methods},
  year          = {2025},
  eprint        = {2505.06471},
  archivePrefix = {arXiv}
}

@inproceedings{masum2025quantum,
  author   = {Masum, Abu Kaisar Mohammad and Shoushtari Moghadam, Mehran and Kouhalvandi, Lida and Najafi, M. Hassan and Aygun, Sercan},
  title    = {Quantum Image Processing: A Comparative Study of {NEQR} and {FRQI} Encoding Schemes with Hybrid Processing},
  booktitle= {Proceedings of the Great Lakes Symposium on VLSI 2025},
  publisher= {ACM},
  pages    = {575--580},
  year     = {2025},
  DOI      = {10.1145/3716368.3735286}
}

@article{khan2019improved,
  author   = {Khan, Rabia Amin},
  title    = {An improved flexible representation of quantum images},
  journal  = {Quantum Information Processing},
  volume   = {18},
  pages    = {201},
  year     = {2019},
  DOI      = {10.1007/s11128-019-2306-6}
}

@inproceedings{leflexible,
  author    = {Le, Phuc Q. and Dong, Fangyan and Arai, Yoshinori and Hirota, Kaoru},
  title     = {Flexible representation of quantum images and its computational complexity analysis},
  booktitle = {Proceedings of the 25th Fuzzy System Symposium},
  publisher = {Japan Society for Fuzzy Theory and Intelligent Informatics},
  year      = {2009}
}

@article{lu2019multimedia,
  author   = {Lu, Zhongyue and Wang, Xiaoming and Shang, Jianzhong and Luo, Zirong and Sun, Chongfei and Wu, Guoheng},
  title    = {A multimedia image edge extraction algorithm based on flexible representation of quantum},
  journal  = {Multimedia Tools and Applications},
  volume   = {78},
  pages    = {24067--24082},
  year     = {2019},
  DOI      = {10.1007/s11042-019-7173-8}
}

@article{Zhang2013,
  author   = {Zhang, Yi and Lu, Kai and Gao, Yinghui and Wang, Mo},
  title    = {{NEQR}: a novel enhanced quantum representation of digital images},
  journal  = {Quantum Information Processing},
  volume   = {12},
  pages    = {2833--2860},
  year     = {2013},
  DOI      = {10.1007/s11128-013-0567-z}
}

@article{du2022binarization,
  author   = {Du, Shiping and Luo, Kailun and Zhi, Yan and Situ, Haozhen and Zhang, Jin},
  title    = {Binarization of grayscale quantum image denoted with novel enhanced quantum representations},
  journal  = {Results in Physics},
  volume   = {39},
  pages    = {105710},
  year     = {2022},
  DOI      = {10.1016/j.rinp.2022.105710}
}

@misc{draper2000addition,
  author        = {Draper, Thomas G.},
  title         = {Addition on a quantum computer},
  year          = {2000},
  eprint        = {quant-ph/0008033},
  archivePrefix = {arXiv}
}

@article{ruizperez2017quantum,
  author  = {Ruiz-Perez, Lidia and Garcia-Escartin, Juan Carlos},
  title   = {Quantum arithmetic with the quantum {F}ourier transform},
  journal = {Quantum Information Processing},
  volume  = {16},
  number  = {6},
  pages   = {152},
  year    = {2017},
  DOI     = {10.1007/s11128-017-1603-1}
}

@article{vedral1996quantum,
  author  = {Vedral, Vlatko and Barenco, Adriano and Ekert, Artur},
  title   = {Quantum networks for elementary arithmetic operations},
  journal = {Physical Review A},
  volume  = {54},
  pages   = {147--153},
  year    = {1996},
  DOI     = {10.1103/PhysRevA.54.147}
}

@article{barenco1996approximate,
  author  = {Barenco, Adriano and Ekert, Artur and Suominen, Kalle-Antti and T{\"o}rm{\"a}, P{\"a}ivi},
  title   = {Approximate quantum {F}ourier transform and decoherence},
  journal = {Physical Review A},
  volume  = {54},
  pages   = {139--146},
  year    = {1996},
  DOI     = {10.1103/PhysRevA.54.139}
}

@article{temme2017error,
  author  = {Temme, Kristan and Bravyi, Sergey and Gambetta, Jay M.},
  title   = {Error mitigation for short-depth quantum circuits},
  journal = {Physical Review Letters},
  volume  = {119},
  pages   = {180509},
  year    = {2017},
  DOI     = {10.1103/PhysRevLett.119.180509}
}

@article{sang2017novel,
  author   = {Sang, Jianzhi and Wang, Shen and Li, Qiong},
  title    = {A novel quantum representation of color digital images},
  journal  = {Quantum Information Processing},
  volume   = {16},
  pages    = {42},
  year     = {2017},
  DOI      = {10.1007/s11128-016-1463-0}
}

@article{su2021improved,
  author   = {Su, Jie and Guo, Xuchao and Liu, Chengqi and Lu, Shuhan and Li, Lin},
  title    = {An improved novel quantum image representation and its experimental test on {IBM} quantum experience},
  journal  = {Scientific Reports},
  volume   = {11},
  pages    = {13879},
  year     = {2021},
  DOI      = {10.1038/s41598-021-93471-7}
}

@article{majji2023quantum,
  author   = {Majji, Sathwik Reddy and Chalumuri, Avinash and Manoj, B. S.},
  title    = {Quantum Approach to Image Data Encoding and Compression},
  journal  = {IEEE Sensors Letters},
  volume   = {7},
  pages    = {1--4},
  year     = {2023},
  DOI      = {10.1109/lsens.2023.3239749}
}

@article{easom2022efficient,
  author   = {Easom-McCaldin, Philip and Bouridane, Ahmed and Belatreche, Ammar and Jiang, Richard and Al-Maadeed, Somaya},
  title    = {Efficient Quantum Image Classification Using Single Qubit Encoding},
  journal  = {IEEE Transactions on Neural Networks and Learning Systems},
  volume   = {35},
  pages    = {1472--1486},
  year     = {2024},
  DOI      = {10.1109/tnnls.2022.3179354}
}

@article{grigoryan2020new,
  author   = {Grigoryan, Artyom M. and Agaian, Sos S.},
  title    = {New look on quantum representation of images: {Fourier} transform representation},
  journal  = {Quantum Information Processing},
  volume   = {19},
  pages    = {148},
  year     = {2020},
  DOI      = {10.1007/s11128-020-02643-3}
}

@article{alwan2025multilayered,
  author   = {Alwan, Nawres A. and Obaiys, Suzan J. and Al-Saidi, Nadia M. G. and Noor, Nurul Fazmidar Binti Mohd and Karaca, Yeliz},
  title    = { Multilayered quantum computing and simulation system for enhanced image representation of {HSI} based {Fourier} transform and adjacency matrix},
  journal  = {Scientific Reports},
  volume   = {15},
  pages    = {40286},
  year     = {2025},
  DOI      = {10.1038/s41598-025-24168-4}
}

@misc{Wereszczynski2020,
  author        = {Wereszczy{\'n}ski, Kamil and Michalczuk, Agnieszka and P{\k{e}}szor, Damian and Paszkuta, Marcin and Cyran, Krzysztof and Pola{\'n}ski, Andrzej},
  title         = {Cosine series quantum sampling method with applications in signal and image processing},
  year          = {2020},
  eprint        = {2011.12738},
  archivePrefix = {arXiv}
}

@incollection{grigoryan2025image,
  author    = {Grigoryan, Artyom M. and Agaian, Sos S.},
  title     = {Image Representation on the Unit Circle and {MQFTR}},
  booktitle = {Quantum Image Processing in Practice: A Mathematical Toolbox},
  publisher = {Wiley},
  pages     = {147--160},
  year      = {2025},
  DOI       = {10.1002/9781394265183.ch8}
}

@article{Werner2023DataLoss,
  author   = {Werner, Krzysztof and Kordasz, Michał and Michalczuk, Agnieszka and Cyran, Krzysztof},
  title    = {Data loss in quantum image representation methods},
  journal  = {Procedia Computer Science},
  volume   = {225},
  pages    = {4354--4363},
  year     = {2023},
  DOI      = {10.1016/j.procs.2023.10.432}
}

@article{Haque2023EFRQI,
  author   = {Haque, Md Ershadul and Paul, Manoranjan and Ulhaq, Anwaar and Debnath, Tanmoy},
  title    = {Advanced quantum image representation and compression using a {DCT}-{EFRQI} approach},
  journal  = {Scientific Reports},
  volume   = {13},
  pages    = {4129},
  year     = {2023},
  DOI      = {10.1038/s41598-023-30575-2}
}

@book{NielsenChuang,
  author    = {Nielsen, Michael A. and Chuang, Isaac L.},
  title     = {Quantum Computation and Quantum Information},
  edition   = {10th anniversary},
  publisher = {Cambridge University Press},
  address   = {Cambridge},
  year      = {2010},
  DOI       = {10.1017/CBO9780511976667}
}

@article{weinstein2001implementation,
  author   = {Weinstein, Y. S. and Pravia, M. A. and Fortunato, E. M. and Lloyd, S. and Cory, D. G.},
  title    = {Implementation of the Quantum {Fourier} Transform},
  journal  = {Physical Review Letters},
  volume   = {86},
  pages    = {1889--1891},
  year     = {2001},
  DOI      = {10.1103/physrevlett.86.1889}
}

@article{camps2021quantum,
  author   = {Camps, Daan and Van Beeumen, Roel and Yang, Chao},
  title    = {Quantum {Fourier} transform revisited},
  journal  = {Numerical Linear Algebra with Applications},
  volume   = {28},
  pages    = {e2331},
  year     = {2021},
  DOI      = {10.1002/nla.2331}
}

@article{yao2017quantum,
  author   = {Yao, Xi-Wei and Wang, Hengyan and Liao, Zeyang and Chen, Ming-Cheng and Pan, Jian and Li, Jun and Zhang, Kechao and Lin, Xingcheng and Wang, Zhehui and Luo, Zhihuang and Zheng, Wenqiang and Li, Jianzhong and Zhao, Meisheng and Peng, Xinhua and Suter, Dieter},
  title    = {Quantum Image Processing and Its Application to Edge Detection: Theory and Experiment},
  journal  = {Physical Review X},
  volume   = {7},
  pages    = {031041},
  year     = {2017},
  DOI      = {10.1103/physrevx.7.031041}
}

@article{llorens2025quantum,
  author   = {Llorens, Santiago and González, Walther and Sentís, Gael and Calsamiglia, John and Muñoz-Tapia, Ramon and Bagan, Emili},
  title    = {Quantum Edge Detection},
  journal  = {Quantum},
  volume   = {9},
  pages    = {1687},
  year     = {2025},
  DOI      = {10.22331/q-2025-04-03-1687}
}

@article{zhou2020metasurface,
  author   = {Zhou, Junxiao and Liu, Shikai and Qian, Haoliang and Li, Yinhai and Luo, Hailu and Wen, Shuangchun and Zhou, Zhiyuan and Guo, Guangcan and Shi, Baosen and Liu, Zhaowei},
  title    = {Metasurface enabled quantum edge detection},
  journal  = {Science Advances},
  volume   = {6},
  pages    = {eabc4385},
  year     = {2020},
  DOI      = {10.1126/sciadv.abc4385}
}

@article{barenco1995elementary,
  author   = {Barenco, Adriano and Bennett, Charles H. and Cleve, Richard and DiVincenzo, David P. and Margolus, Norman and Shor, Peter and Sleator, Tycho and Smolin, John A. and Weinfurter, Harald},
  title    = {Elementary gates for quantum computation},
  journal  = {Physical Review A},
  volume   = {52},
  pages    = {3457--3467},
  year     = {1995},
  DOI      = {10.1103/physreva.52.3457}
}

@article{chang2022improving,
  author   = {Chang, Chia Cheng and McElvain, Kenneth S. and Rrapaj, Ermal and Wu, Yantao},
  title    = {Improving {Schr}ödinger Equation Implementations with {Gray} Code for Adiabatic Quantum Computers},
  journal  = {PRX Quantum},
  volume   = {3},
  pages    = {020356},
  year     = {2022},
  DOI      = {10.1103/prxquantum.3.020356}
}

@article{spagnolini19932,
  author   = {Spagnolini, Umberto},
  title    = {2-D phase unwrapping and phase aliasing},
  journal  = {Geophysics},
  volume   = {58},
  pages    = {1324--1334},
  year     = {1993},
  DOI      = {10.1190/1.1443515}
}

@article{zhu2022anti,
  author   = {Zhu, Huijie and Guo, Hongwei},
  title    = {Anti-aliasing phase reconstruction via a non-uniform phase-shifting technique},
  journal  = {Optics Express},
  volume   = {30},
  pages    = {3835},
  year     = {2022},
  DOI      = {10.1364/oe.449656}
}

@article{breitenbach1999against,
  author   = {Breitenbach, Arvid},
  title    = {Against spectral leakage},
  journal  = {Measurement},
  volume   = {25},
  pages    = {135--142},
  year     = {1999},
  DOI      = {10.1016/s0263-2241(98)00074-8}
}

@article{zhou2017quantum,
  author   = {Zhou, S. S. and Loke, T. and Izaac, J. A. and Wang, J. B.},
  title    = {Quantum {Fourier} transform in computational basis},
  journal  = {Quantum Information Processing},
  volume   = {16},
  pages    = {82},
  year     = {2017},
  DOI      = {10.1007/s11128-017-1515-0}
}

@article{li2024iterative,
  author   = {Li, Junxu},
  title    = {Iterative method to improve the precision of the quantum-phase-estimation algorithm},
  journal  = {Physical Review A},
  volume   = {109},
  pages    = {032606},
  year     = {2024},
  DOI      = {10.1103/physreva.109.032606}
}

@article{amankwah2022quantum,
  author   = {Amankwah, Mercy G. and Camps, Daan and Bethel, E. Wes and Van Beeumen, Roel and Perciano, Talita},
  title    = {Quantum pixel representations and compression for {N-dimensional} images},
  journal  = {Scientific Reports},
  volume   = {12},
  pages    = {7712},
  year     = {2022},
  DOI      = {10.1038/s41598-022-11024-y}
}

@article{mottonen2004quantum,
  author   = {Möttönen, Mikko and Vartiainen, Juha J. and Bergholm, Ville and Salomaa, Martti M.},
  title    = {Quantum Circuits for General Multiqubit Gates},
  journal  = {Physical Review Letters},
  volume   = {93},
  pages    = {130502},
  year     = {2004},
  DOI      = {10.1103/physrevlett.93.130502}
}

@article{bullock2004asymptotically,
  author   = {Bullock, S.S. and Markov, I.L.},
  title    = {Asymptotically optimal circuits for arbitrary n-qubit diagonal computations},
  journal  = {Quantum Information and Computation},
  volume   = {4},
  pages    = {27--47},
  year     = {2004},
  DOI      = {10.26421/qic4.1-3}
}

@article{z33f-l4jg,
  title = {Quantum image encodings for noisy intermediate-scale quantum devices: A comprehensive performance and noise-resilience analysis},
  author = {Mart\'{\i}nez-Guerrero, Emmanuel and Sun, Guo-Hua and Dong, Shi-Hai},
  journal = {Phys. Rev. A},
  volume = {113},
  issue = {6},
  pages = {062456},
  numpages = {27},
  year = {2026},
  month = {Jun},
  publisher = {American Physical Society},
  doi = {10.1103/z33f-l4jg}
}

@article{welch2014efficient,
  author   = {Welch, Jonathan and Greenbaum, Daniel and Mostame, Sarah and Aspuru-Guzik, Alan},
  title    = {Efficient quantum circuits for diagonal unitaries without ancillas},
  journal  = {New Journal of Physics},
  volume   = {16},
  pages    = {033040},
  year     = {2014},
  DOI      = {10.1088/1367-2630/16/3/033040}
}

@article{shende2006synthesis,
  author   = {Shende, V.V. and Bullock, S.S. and Markov, I.L.},
  title    = {Synthesis of quantum-logic circuits},
  journal  = {IEEE Transactions on Computer-Aided Design of Integrated Circuits and Systems},
  volume   = {25},
  pages    = {1000--1010},
  year     = {2006},
  DOI      = {10.1109/tcad.2005.855930}
}

@article{dasilva2022linear,
  author   = {da Silva, Adenilton J. and Park, Daniel K.},
  title    = {Linear-depth quantum circuits for multiqubit controlled gates},
  journal  = {Physical Review A},
  volume   = {106},
  pages    = {042602},
  year     = {2022},
  DOI      = {10.1103/physreva.106.042602}
}

@article{cleve1998quantum,
  author   = {Cleve, R. and Ekert, A. and Macchiavello, C. and Mosca, M.},
  title    = {Quantum algorithms revisited},
  journal  = {Proceedings of the Royal Society of London. Series A: Mathematical, Physical and Engineering Sciences},
  volume   = {454},
  pages    = {339--354},
  year     = {1998},
  DOI      = {10.1098/rspa.1998.0164}
}

@misc{kitaev1995quantum,
  author        = {Kitaev, A. Yu.},
  title         = {Quantum measurements and the {Abelian} Stabilizer Problem},
  year          = {1995},
  eprint        = {quant-ph/9511026},
  archivePrefix = {arXiv}
}

@article{aaronson2015read,
  author   = {Aaronson, Scott},
  title    = {Read the fine print},
  journal  = {Nature Physics},
  volume   = {11},
  pages    = {291--293},
  year     = {2015},
  DOI      = {10.1038/nphys3272}
}

@article{harris1978use,
  author   = {Harris, F.J.},
  title    = {On the use of windows for harmonic analysis with the discrete {Fourier} transform},
  journal  = {Proceedings of the IEEE},
  volume   = {66},
  pages    = {51--83},
  year     = {1978},
  DOI      = {10.1109/proc.1978.10837}
}

@article{zhang2015qsobel,
  author   = {Zhang, Yi and Lu, Kai and Gao, YingHui},
  title    = {{QSobel}: A novel quantum image edge extraction algorithm},
  journal  = {Science China Information Sciences},
  volume   = {58},
  pages    = {1--13},
  year     = {2015},
  DOI      = {10.1007/s11432-014-5158-9}
}

@article{havlicek2019supervised,
  author   = {Havlíček, Vojtěch and Córcoles, Antonio D. and Temme, Kristan and Harrow, Aram W. and Kandala, Abhinav and Chow, Jerry M. and Gambetta, Jay M.},
  title    = {Supervised learning with quantum-enhanced feature spaces},
  journal  = {Nature},
  volume   = {567},
  pages    = {209--212},
  year     = {2019},
  DOI      = {10.1038/s41586-019-0980-2}
}

@book{gonzalez2018digital,
  author    = {Gonzalez, Rafael C. and Woods, Richard E.},
  title     = {Digital Image Processing},
  edition   = {4th},
  publisher = {Pearson},
  year      = {2018}
}

@misc{du2018synthesis,
  author        = {Du, Shiping and Qiu, Daowen and Gruska, Jozef and Mateus, Paulo and others},
  title         = {Synthesis of Quantum Images Using Phase Rotation},
  year          = {2018},
  eprint        = {1811.05170},
  archivePrefix = {arXiv}
}

@misc{herrero2026spectral,
  author        = {Herrero G{\'o}mez, Pablo and Jimeno Morenilla, Antonio and others},
  title         = {Spectral Phase Encoding for Quantum Kernel Methods},
  year          = {2026},
  eprint        = {2602.19644},
  archivePrefix = {arXiv}
}

@article{zhai2025quantum,
  author   = {Zhai, Xinliang and Xiao, Tailong and Huang, Jingzheng and Fan, Jianping and Zeng, Guihua},
  title    = {Quantum neural compressive sensing for ghost imaging},
  journal  = {Physical Review Applied},
  volume   = {23},
  pages    = {014018},
  year     = {2025},
  DOI      = {10.1103/physrevapplied.23.014018}
}

@article{xiao2023practical,
  author   = {Xiao, Tailong and Zhai, Xinliang and Wu, Xiaoyan and Fan, Jianping and Zeng, Guihua},
  title    = {Practical advantage of quantum machine learning in ghost imaging},
  journal  = {Communications Physics},
  volume   = {6},
  pages    = {171},
  year     = {2023},
  DOI      = {10.1038/s42005-023-01290-1}
}

@article{akinola2024robust,
  author   = {Akinola, Taoreed A. and Li, Xiangfang and Wilkins, Richard and Obiomon, Pamela H. and Qian, Lijun},
  title    = {Robust Inverse Quantum {Fourier} Transform Inspired Algorithm for Unsupervised Image Segmentation},
  journal  = {IEEE Access},
  volume   = {12},
  pages    = {99029--99044},
  year     = {2024},
  DOI      = {10.1109/access.2024.3429413}
}

\end{document}